 \documentclass[a4paper,preprint,pra,superscriptaddress,showpacs,floatfix,nofootinbib]{revtex4}

\usepackage{dcolumn}
\usepackage{amsmath}
\usepackage{soul}
\usepackage{adjustbox}
\usepackage{ulem}
\usepackage{array}
\usepackage{hyperref}
\usepackage{lipsum}
\usepackage{booktabs}                     

\usepackage[english]{babel}
\usepackage[T1]{fontenc}
\usepackage{amssymb}
\usepackage{graphicx}
\graphicspath{.}
\usepackage{bm}
\usepackage{amsmath}
\usepackage{natbib}
\usepackage[mathcal]{eucal}
\usepackage{numprint}
\usepackage{lscape}
\usepackage{url}
\usepackage{tikz}
\usetikzlibrary{shapes}
\usetikzlibrary{decorations}
\usetikzlibrary{arrows}
\usetikzlibrary{matrix}
\usepackage{appendix}
\usepackage{footmisc}
\usepackage{comment}
\usepackage{tabularx}
\usepackage{float}
\usepackage{csquotes}
\usepackage{color}
\usepackage{xcolor}

\newcommand {\nc} {\newcommand}
\nc {\eoln} [1] {\label{#1} \\}
\nc {\eol} {\nonumber \\}
\nc {\rref} [1] {(\ref{#1})}
\nc {\Eq} [1] {Eq.~(\ref{#1})}
\nc {\Ref} [1] {Ref.~\cite{#1}}
\nc {\la} {\mbox{$\langle$}}
\nc {\ra} {\mbox{$\rangle$}}
\nc {\dem} {\mbox{$\frac{1}{2}$}}
\nc {\cP} {\mathcal{P}}
\nc {\cN} {\mathcal{N}}
\nc {\ve} [1] {\mbox{\boldmath $#1$}}
\nc {\arrow} [2] {\mbox{$\mathop{\rightarrow}\limits_{#1 \rightarrow #2}$}}
\nc {\red}[1] {\textcolor{red}{#1}}
\nc {\mc}[3] {\multicolumn{#1}{#2}{#3}}
\nc {\dd}{\, \mathrm{d}}
\nc {\bs}[1]{\boldsymbol{#1}}
\nc {\ket}[1]{\vert #1 \rangle}
\nc {\bra}[1]{\langle #1 \vert}
\nc {\abs}[1]{\vert #1 \vert}
\nc {\avg}[1]{\langle #1 \rangle}
\nc {\braket}[2]{\langle #1 \vphantom{#2} \vert #2 \vphantom{#1} \rangle}

\def\pu1#1{{\color{purple}{#1}}}  
\def\te1#1{{\color{teal}{#1}}}  
\def\li1#1{{\color{lime}{#1}}}  
\def\pi1#1{{\color{pink}{#1}}}  
\def\ye1#1{{\color{yellow}{#1}}}  

\makeatletter
\renewcommand\@makefntext[1]%
     {\noindent\makebox[0pt][r]{\textsuperscript{\@thefnmark}\,}#1}
\makeatother

\begin{document}
     

\title{\textit{Ab initio} electronic factors of the $A$ and $B$ hyperfine structure constants for the
  $5s^25p6s \; ^{1,3}\! P^{\rm o}_{1} $
  states in Sn~I}
 %
\author{Asimina Papoulia\footnotemark[4]}
\email[]{asimina.papoulia@mau.se}
\affiliation{Department of Materials Science and Applied Mathematics, Malm\"o University, SE-20506 Malm\"o, Sweden}
\affiliation{Division of Mathematical Physics, Department of Physics, Lund University, SE-22100 Lund, Sweden}
\author{Sacha Schiffmann\footnotemark[4]}
\email[]{saschiff@ulb.ac.be}
\affiliation{Division of Mathematical Physics, Department of Physics, Lund University, SE-22100 Lund, Sweden}
\affiliation{Spectroscopy, Quantum Chemistry and Atmospheric Remote Sensing (SQUARES), CP160/09, Universit\'e libre de Bruxelles (ULB), 1050 Brussels, Belgium}
\renewcommand{\thefootnote}{\fnsymbol{footnote}}
\footnotetext[4]{Contributed equally to this work.}
\author{Jacek Biero{\'n}}
\affiliation{Instytut Fizyki Teoretycznej,
             Uniwersytet Jagiello{\'n}ski,
             ul.~prof.~Stanis{\l}awa {\L}ojasiewicza 11,
             Krak{\'o}w, Poland}
%
%
\author{Gediminas Gaigalas}
\affiliation{Institute of Theoretical Physics and Astronomy, Vilnius  University, Saul\.etekio av.~3, LT-10222, Vilnius, Lithuania}
\author{Michel Godefroid}
\affiliation{Spectroscopy, Quantum Chemistry and Atmospheric Remote Sensing (SQUARES), CP160/09, Universit\'e libre de Bruxelles (ULB), 1050 Brussels, Belgium}
\author{Zolt\'an Harman}
\affiliation{Max Planck Institute for Nuclear Physics, Saupfercheckweg 1, 69117 Heidelberg, Germany}
\author{Per~J\"onsson}
\affiliation{Department of Materials Science and Applied Mathematics, Malm\"o University, SE-20506 Malm\"o, Sweden}
\author{Natalia S. Oreshkina}
\email[]{natalia.oreshkina@mpi-hd.mpg.de}
\affiliation{Max Planck Institute for Nuclear Physics, Saupfercheckweg 1, 69117 Heidelberg, Germany}
\author{Pekka Pyykk\"o}
\affiliation{Department of Chemistry, University of Helsinki, PO Box 55 (A. I. Virtasen aukio 1), FIN-00014 Helsinki, Finland}
\author{Ilya I. Tupitsyn}
\affiliation{Department of Physics, St. Petersburg State University, 198504 St. Petersburg, Russia}

\date{\today}

\begin{abstract}
Large-scale \textit{ab initio} calculations of the electric field gradient, which constitutes the electronic contribution to the electric quadrupole hyperfine constant $B$, were performed for the $5s^25p6s$ $^{1,3}\!P^{\rm o}_1$ excited states of tin, using three independent computational strategies of the variational multiconfiguration Dirac-Hartree-Fock method and a fourth approach based on the configuration interaction Dirac-Fock-Sturm theory. For the $5s^25p6s$ $^{1}\!P^{\rm o}_1$ state, the final value of $B/Q =703(50)$~MHz/b  differs by $0.4\%$ from the one recently used by Yordanov {\it et~al.} [Communications Physics {\bf 3}, 107 (2020)]
to extract the nuclear quadrupole moments, $Q$, for tin isotopes in the range $^{(117-131)}$Sn from  collinear laser spectroscopy measurements.
Efforts were made to provide a realistic theoretical uncertainty for the final $B/Q$ value of the $5s^25p6s\,^{1}\!P^{\rm o}_1$ state  based on statistical principles and on correlation with the 
magnetic dipole hyperfine constant~$A$. 
%
\end{abstract}

%
%

\maketitle


%
%




\clearpage

\section{Introduction}
\label{section.Introduction}
Nuclear quadrupole moments, $Q$, are important characteristics of nuclei that provide a measure of the deviation of the nuclear charge distribution from a spherical shape. They can be determined from nuclear, atomic, molecular or solid-state spectroscopies,
such as high-resolution laser spectroscopy~\cite{Yordanov:20}, muonic or pionic x-ray spectroscopy~\cite{Antognini2020}, 
Nuclear Magnetic  Resonance (NMR)~\cite{CohRei:57a,Schwerdtfeger:04}, Nuclear Quadrupole Resonance (NQR)~\cite{DasHah:58a,Luc:69a}, M{\"o}ssbauer measurements~\cite{DicBer:2005a,CheYan:2007a} or Perturbed Angular Correlation (PAC) of nuclei passing thin foils~\cite{HaaShi:73a,Haas2017}. 
Most of these techniques involve the knowledge of the electric field gradient (EFG) due to the charges of the constituting environment of the considered nuclei.
Three compilations of available $Q$-values
are provided by Raghavan~\cite{Rag:89a}, Stone~\cite{NJStone:2016},
and Pyykk{\"o}~\cite{Pyykko:18aa}.

In this work, we focus on tin, with an atomic number $Z=50$. The proton shells at this magic number are
closed, but the incomplete neutron shells can still induce a
$Q$ with quadratic dependence on the neutron number $N$, which will not 
become magic until the known $^{132}$Sn isotope, with $N=82$ and nuclear spin $I=0$. 
The nuclear trends of $Q$ and the charge radii among ten isotopes in the range $^{117-131}$Sn, which is below the doubly magic isotope, have just been published by Yordanov {\it et al.}~\cite{Yordanov:20}.
Continuous, linear and quadratic trends 
were found as functions of the mass number for cadmium 
and tin isotopes, respectively.
The $Q$(Sn)-values published in Ref.~\cite{Yordanov:20} were
based on measured atomic hyperfine structures for odd-$N$ isotopes and were obtained by combining the calculated EFG in the [Pd]$5s^25p6s~{}^1\!P_1^{\rm o}$ state of the neutral atom with the measured electric quadrupole (E2) hyperfine coupling constant, $B$, for each isotope. 
(Note that [Pd] is used, for brevity, to indicate the palladium-like core and will be omitted in the following.)
The relative accuracy of the calculated EFG value in the $5s^25p6s~{}^1\!P_1^{\rm o}$ state is of the order of 7\%.
The accuracy of the measured $B$-values varies, depending on the isotope,
between 1.5\% for $^{131}$Sn and 33\% for $^{125}$Sn, as reflected in Table~1 of Ref.~\cite{Yordanov:20}. As a result, the accuracy of the evaluated \enquote{atomic} $Q$(Sn)-values ranges between 7\% and 34\%.

No \enquote{pionic} nor \enquote{muonic} $Q$(Sn)-values are available.
There exists a quoted value of $-$132(1)~mb
for the $I$~=~3/2,~24-keV M{\"o}ssbauer state of $^{119}$Sn
by Barone \textit{et~al.}~\cite{Barone:08}, using data from 34 solids.
A closely similar |$Q$| value of 128(7) mb was reported by
Svane~\textit{et~al.}~\cite{Svaetal:97a}.
Moreover, the same 24-keV, $I=3/2$ state of $^{119}$Sn was observed by Dimmling \textit{et~al.}~\cite{Dimmling:75} using PAC, together with other tin isotopes embedded in elemental Cd, Sn, or Sb. Taking their $B$-measurements at face value
and renormalizing to the $-$132(1)~mb value of Ref.~\cite{Barone:08} by using the experimental ratio of 2.2(2) from Beloserski \textit{et~al.}~\cite{Beletal:72a}, we obtain 
\begin{equation} 
\label{eq:renormQ}
Q(^{119}{\rm Sn},~I=11/2,~90~{\rm keV}) =  -132(1) \frac{Q(11/2)}{Q(3/2) } = -132(1) \cdot 2.2(2) = -290~\mbox{mb}.
\end{equation}
In fact, such a value was already used in the review by Stone~\cite{NJStone:2016}.
It can be compared with the $Q(^{119}{\rm Sn})=-$175(4)(12) mb in Yordanov \textit{et al.}~\cite{Yordanov:20}, where a direct atomic measurement was combined 
with a calculation of the EFG. Thus, some coupling between the PAC and M{\"o}ssbauer $Q$-values and the present \enquote{atomic} ones seems possible. We, finally, note that Ref.~\cite{NJStone:2016} gives some $Q$-values for the 
even-$N$ isotopes within the range $^{110-124}{\rm Sn}$, and also 
$^{130}{\rm Sn}$, in various nuclear states. Yet, none of the observed hyperfine interaction constants $B$ that were used for extracting these $Q$(Sn)-values have very high accuracy.

The aim of the present paper is to report the details of the atomic calculations of the EFG values for the $5s^25p6s$ $^{1,3}\!P^{\rm o}_1$ excited states in neutral tin. The EFG constitutes the electronic part of the E2 hyperfine constant $B$. This electronic contribution is, in Sec.~\ref{subsection.hfs_theory}, defined as $B/Q\equiv B_{\text{el}}$. A well-grounded estimate of the theoretical uncertainties is provided by evaluating the computed values of both E2 and magnetic dipole (M1) hyperfine structures.

Following the multiconfiguration Dirac-Hartree-Fock (MCDHF) method~\cite{GrantBook2007,CFFreview:2016} described in Sec.~\ref{subsection.MCDHF-RCI}, three different computational schemes for optimizing the Dirac one-electron radial functions were employed. The first optimization strategy, denoted S-MR-MCDHF, is based on configuration state function (CSF) expansions that were built from single (S) electron substitutions from a set of multi-reference (MR) configurations, while the second approach, denoted SrD-SR-MCDHF, is based on CSF expansions produced by S and restricted double (rD) substitutions from a single-reference (SR) configuration. Finally, the third optimization scheme, named SrD-MR-MCDHF, is similar to the second approach, regarding the electron substitutions that build the CSF space, but it is instead applied to a selected MR. 
All three approaches include higher-order electron substitutions in the subsequent relativistic configuration interaction steps. 
The sets of calculations based on the above-mentioned optimization schemes are, respectively, discussed in Secs.~\ref{section.Heidelberg},~\ref{section.SRcalculations}, and~\ref{section.MRcalculations}. A fourth independent set of calculations was performed using the configuration interaction Dirac-Fock-Sturmian (CI-DFS) method~\cite{Tup2003OS,Tup2003PRA,Tup2005,SoriaOrts2006} described in Sec.~\ref{subsection.CIDFS}, and the respective values of the computed electronic hyperfine factors are reported in Sec.~\ref{section.CIDFS_calculations}. 

\renewcommand{\thefootnote}{\arabic{footnote}}
The results from all four different approaches are combined to provide theoretical uncertainties  in Sec.~\ref{section.ErrorBudget}. Note that the latest of the four independent calculations that are presented here was carried out after the work by Yordanov {\it et al.}~\cite{Yordanov:20} had been sent to the printer and represent new, previously unpublished, results. That being so, in the current paper, the final $B/Q\equiv B_{\text{el}}$ value for the $5s^25p6s~{}^1\!P_1^o$ state is slightly shifted from 706(50)~MHz/b~\cite{Yordanov:20} to 703(50)~MHz/b\footnotemark[1].
\footnotetext[1]{Although the electronic contribution $B/Q$ is proportional to the computed EFG value (see also Sec.~\ref{subsection.hfs_theory}), in Ref.~\cite{Yordanov:20}, the quantities EFG and $B/Q$ are used interchangeably.}


\section{Theory}
\label{section.Theory}
\subsection{MCDHF-RCI multiconfiguration methods}
\label{subsection.MCDHF-RCI}
The principles of the MCDHF method are fully discussed in, e.g., the book by Grant~\cite{GrantBook2007} and the review article by Froese Fischer~\textit{et~al.}~\cite{CFFreview:2016}. With this section, we provide the reader with a short introduction of the main concepts.

In the relativistic framework, the MCDHF 
method describes an atomic state function (ASF), $\Psi ( \gamma \Pi J M)$, as an expansion over a set of $jj$-coupled relativistic CSFs, $\Phi_{\mu}(\gamma_\mu  \Pi JM)$, characterized by the parity~$\Pi$, the total electronic angular momentum $J$, and the projection quantum numbers $M$,  i.e.,
\begin{equation}
\label{eq:jj_ASF}
  \Psi ( \gamma \Pi \\J M) = \sum_{\mu=1}^{N_{\text{CSF}}} c_\mu \Phi_{\mu}(\gamma_\mu  \Pi \\J M), \mbox{\ where }
\sum_{\mu=1}^{N_{\text{CSF}}} c_\mu^2 = 1.
\end{equation}
In the expression above, $\gamma_\mu$ represents the configuration, the angular momentum coupling tree, and other quantum numbers that are necessary to uniquely describe each CSF. The CSFs are built on one-electron Dirac spinors that have the general form:
\vspace{0.1cm}
\begin{equation}
\label{eq:R_orbitals_kappa}
\psi_{n  \kappa  m} 
(r, \theta, \varphi)
 = \frac{1}{r}
 \left( 
\begin{array}{c}
P_{n  \kappa}(r) \; {\Omega}_{\kappa m}(\theta, \varphi) \\
\mbox{i} \; Q_{n  \kappa}(r) \; {\Omega}_{-\kappa m}(\theta, \varphi)  
\end{array} 
\right),
\end{equation}
where $\Omega_{\pm \kappa m}(\theta,\varphi)$ are the two-component spin-angular functions and $\{ P_{n \kappa} (r)$, $Q_{n \kappa} (r)\}$ are, respectively, the radial functions of the so-called large and small components with principal quantum number $n$ and relativistic angular momentum quantum number $\kappa$ defined as
\begin{equation}
\label{eq:def2_kappa}
\kappa = \left\{
\begin{array}{cl}
 -(l+1)  &\hspace{3 mm} \mbox{for} \hspace{3 mm}  j = l + 1/2 \hspace{3 mm} (\kappa~\mbox{negative})\\
   +l    &\hspace{3 mm} \mbox{for} \hspace{3 mm}  j = l - 1/2 \hspace{3 mm} (\kappa~\mbox{positive})\\
\end{array}\right. \; ,
\end{equation}
where $l$ is the orbital angular momentum quantum number. The quantum number $m$ describing the Dirac spinors \eqref{eq:R_orbitals_kappa} is the projection of the total angular momentum~$j$.

To derive the MCDHF equations, the variational principle is applied on the functional:
\begin{equation}
\label{eq:functional_MCDHF}
 {\cal F}(\{c\},\{P\},\{Q\}; \gamma \Pi J) \equiv 
     \langle \Psi \vert {\cal H}_{\rm DC}\vert \Psi \rangle + \sum_{ab} \delta_{\kappa_a \kappa_b} 
     \lambda_{ab} \;  {\cal C}_{ab}, 
\end{equation}
where the indices $a$ and $b$, respectively, represent the orbitals $n_a \kappa_a$ and $n_b \kappa_b$, which are used to construct the ASF of Eq.~(\ref{eq:jj_ASF}). The energy functional is estimated from the expectation value of the $N_e$-electron Dirac-Coulomb Hamiltonian
\begin{equation}
\label{eq:DC}
{\cal H}_{\rm DC}= \sum_{i=1}^{N_e} h_{\rm D}(i) + \! \sum_{j>i=1}^{N_e} \frac{1}{r_{ij}} =
\sum_{i=1}^{N_e} \left[ c \; \boldsymbol{\alpha}_i \cdot {\bf p}_i + c^2 (\beta_i-1) + V_{\rm nuc}(r_i)  \right]  + \!\sum_{j>i=1}^{N_e} \frac{1}{r_{ij}},
\end{equation}
where $V_{\text{nuc}}(r_i)$ is the potential from an extended nuclear charge distribution, $\bm{\alpha}$ and $\beta$ are the $4 \times 4$ Dirac matrices, $c$ is the speed of light in atomic units, and $\textbf{p} \equiv -\rm{i}\boldsymbol{\nabla}$ is the electron momentum operator. Lagrange multipliers $\lambda_{ab}$ are introduced for constraining the variations $(\delta P_{n \kappa}, \delta Q_{n \kappa})$ to 
satisfy the orthonormality of the one-electron function, i.e.,
\begin{equation}
\label{eq:rel_radial_ovlp}
{\cal C}_{ab} \equiv \int_0^\infty \left[ P_a(r)P_b(r)+Q_a(r)Q_b(r) \right]dr - \delta_{ n_a n_b} =0 ,
\end{equation}
and to also ensure the orthonormality of the CSFs. 
The resulting coupled integro-differential equations have the form
\vspace{0.1cm}
\begin{eqnarray}
\label{eq:DHF-operator}
w_a \left[ \begin{array}{c c} 
V(a;r) & -c\left(\frac{d\;}{dr} -\frac{\kappa_a}{r}\right)\\
c\left(\frac{d\;}{dr} +\frac{\kappa_a}{r}\right) & V(a;r) -2c^2 \end{array}\right]\!\left[\begin{array}{c} 
P_{a}(r) \\
Q_{a}(r)\end{array}\right]\!=
\sum_b \epsilon_{ab} \;  \delta_{\kappa_a \kappa_b} \left[
\begin{array}{c}
\!P_{b}(r) \\
\!Q_{b}(r)\end{array}\right] \; , 
\end{eqnarray}
where $w_a$ is the generalized occupation number of the orbital $a$ and $V(a;r) = V_{\rm nuc}(r) + Y(a;r) + \bar{X}(a;r)$ is the average and central field MCDHF potential built from the nuclear, direct, and exchange contributions that arise from both diagonal and off-diagonal 
$\langle \Phi_{\mu} \vert {\cal H}_{\rm DC}  \vert \Phi_{\nu} \rangle$ matrix elements.  
In each $\kappa$-space, Lagrange-related energy parameters $\epsilon_{ab} \equiv \epsilon_{n_a n_b}$ are introduced to impose the orthonormality constraints~(\ref{eq:rel_radial_ovlp}) in the variational process. 

The resulting coupled radial equations are solved iteratively in the self-consistent field (SCF) procedure. Once the radial functions have been determined, a relativistic configuration interaction (RCI) calculation is performed over the set of configuration states to determine the expansion coefficients $c_\mu$ for building the next-iteration potentials. The SCF and RCI coupled processes are repeated until convergence of the total wave function~(\ref{eq:jj_ASF}) and corresponding energy, $\langle \Psi \vert {\cal H}_{\rm DC}  \vert \Psi \rangle$, is reached.

As mentioned in the introduction, these MCDHF  calculations provide the one-electron orbital basis that can be used to investigate the effect of higher-order electron substitutions through RCI calculations that use  larger CSF spaces. At this subsequent RCI step, the transverse photon interaction, which reduces to the Breit interaction at the low-frequency limit, and the leading quantum electrodynamic (QED) corrections are added to the Dirac-Coulomb Hamiltonian (see Refs. \cite{GraPyp:76a,Zhaetal:2020a} for more details).

The MCDHF method is implemented in the {\sc Grasp2K}~\cite{grasp2K:2013} and {\sc Grasp2018}~\cite{GRASP2018} computer packages.
The description of the numerical methods, virtual orbital sets,
electron substitutions, and other details of the computations
can be found in Refs.~\cite{grasp2K:2013,CFFreview:2016,
BieronBeF1999,BieronAu2009,JonssonBieron2010,Bieron:e-N:2015}.
The wave function representation in $jj$-coupling is transformed to an approximate
representation in $LSJ$-coupling,
using the methods and program developed by Gaigalas and
co-workers~\cite{Transformation,JJ2LSJ}.

%

\subsection{CI-DFS method}
\label{subsection.CIDFS}

The detailed description of the CI-DFS method can be found in Refs.~\cite{Tup2003OS,Tup2003PRA,Tup2005,SoriaOrts2006,Tupitsyn2008}.  We  resume hereafter the underlying theoretical background.

\subsubsection {Dirac-Fock-Sturm orbitals.}

Dirac-Fock-Sturm orbitals of a general type $\varphi_j$ can be obtained as the solutions of the following eigenvalue problem:
\begin{equation}
(h_{\rm D}- \varepsilon) \, \varphi_j = \lambda_j \, W(r) \, \varphi_j \,,
\label{Eq:NO_Sturm1}
\end{equation}
where $h_{\rm D}$ is the one-electron Hamiltonian from Eq.~(\ref{eq:DC}), $W(r)$ is the so-called weight function of the positive sign, $\varepsilon$ is a reference energy, and $\lambda_j$ is the eigenvalue.
Analytic Coulomb Sturmian functions, introduced in Ref.~\cite{Rotenberg_1970}, can be obtained for the  non-relativistic H-like Hamiltonian $h_{\rm D}$, $W(r)=1/r$ and $\varepsilon$ is equal to the energy of the $1s$ state. 
This method of Sturmian-function generation is equivalent to the charge-quantization formalism, i.e., to the replacement $Z~\rightarrow~Z^{\ast}_j=Z+\lambda_j$.
The Coulomb Sturmian basis has been shown to be a valuable tool in non-relativistic atomic physics (see, e.g., Ref.~\cite{Avery_2006}).

In the relativistic case, this method fails, since the Dirac equation has no solution for $Z > 137$.
Therefore, early attempts to define the relativistic Sturmians were based on re-writing the Dirac equation for a H-like ion to the second-order equation.
Later, more suitable constructions based on the Dirac equation were proposed by Drake and Goldman~\cite{Drake_1988}, by Grant~\cite{Grant_1991}~(L-spinors), and by Szmytkowski~\cite{Szmytkowski_1997} (see also references therein).

A much more flexible Sturmian basis can be constructed by solving the Sturm equation~\eqref{Eq:NO_Sturm1} numerically for a general form of the weight function $W(r)$. 
In Refs.~\cite{Tup2003OS, Tup2003PRA}, it was proposed to use the Hartree-Fock or Dirac-Fock (DF) operator, in non-relativistic and relativistic cases, respectively,  and the following weight function:
\begin{equation}
W(r) = \left[ \frac{1- \exp(-(\beta r)^2)}{(\beta r)^2} \right] \,,
\label{Eq:NO_weight}
\end{equation}
where the parameter $\beta$ is chosen to speed up the convergence of Sturmian series. 
In contrast to the function $1/r$, the weight function \eqref{Eq:NO_weight} is regular at the origin. 
One can also see that, for $\lambda_j=0$, the Sturmian function coincides with the reference DF orbital.
Since the weight function $W(r) \to 0$ at $r \to \infty$, all Sturmian functions $\varphi_j$ have the same exponential asymptotic behavior at $ r \to \infty$:
\begin{equation}
\varphi_j(r) \to A_j \, e^{-\sqrt{-2 \, \varepsilon}\, r} \,.
\end{equation}
Therefore, all Sturmian functions have approximately the same radial size, which is determined by the reference energy $\varepsilon$.
It is well known that the Sturmian operator is Hermitian and, in contrast to the Fock operator, it does not contain continuous spectra.
Thus, the set of the Sturmian eigenfunctions forms a discrete and complete orthonormalized basis set of one-electron wave functions with the weight  $W(r)$.

\subsubsection {One-electron basis}

The numerical solution of the Dirac-Hartree-Fock equation in the non-relativistic configuration average ($LS$-average) approximation \cite{LindgrenRosen:74a,Tupitsyn_2018} yields one-electron radial wave functions for the occupied (spectroscopic) and low-lying excited orbitals.
The remaining virtual orbitals with positive energies and all orbitals with negative energies have been obtained by solving the generalized Dirac-Fock-Sturm equation~\eqref{Eq:NO_Sturm1} with the weight function~\eqref{Eq:NO_weight} and DF operator as the one-electron Hamiltonian.
The basis constructed in such a way automatically satisfies the dual kinetic balance condition \cite{Shabaev_2004}.

The next step is the construction of an orthonormalized set of one-electron wave functions by the solution of DF equations in the DFS orbital basis. One-electron wave functions obtained before by the DF method stay intact, whereas the virtual Sturmian orbitals are modified to be eigenfunctions of the DF operator and, therefore, they can be used for the construction of determinants in the configuration interaction (CI) method.

\subsubsection {{The Restricted Active Space concept}}
\label{subsubsubsection.RAS}

In Refs.~\cite{Olsen_1988, Malmqvist2002230}, the Restricted Active Space (RAS) concept for electronic structure calculations has been proposed. All non-relativistic atomic shells are divided into three subspaces, namely, RAS1, RAS2, and RAS3. The subspace RAS1 contains the low-lying occupied orbitals (except for the inactive ones), the subspace RAS2 is built from active occupied and low-lying valence orbitals, and all other obitals form RAS3. 

The non-relativistic configurations are constructed by allowing electron substitutions from one or a few orbitals. From the subspace RAS1, $N_{\rm h}$ substitutions are possible to RAS2 or RAS3, usually at most two. Therefore, the configurations with one or two holes in the RAS1 shells  are included. For RAS2, the substitutions are allowed within RAS2 and to RAS3, with the total number of  substitutions $N_{\rm ex}=2,3, \ldots$.
The number of substitutions to subspace RAS3 is limited to $N_{\rm el}$ electrons, which is usually $\le 2$. Subsequently, from the list of non-relativistic configurations a list of relativistic ones is built, including all relativistic configurations that correspond to a given non-relativistic one.
For each relativistic configuration, the CSF set is constructed as linear combinations of Slater determinants, thus forming a many-electron basis for CI calculations.
The substitutions from RAS1 and to RAS3, which are not expected to contribute much compared to substitutions involving the subspace RAS2, can also be taken into account by perturbation theory (PT)~\cite{Tupitsyn2008}.

%
%
\subsection{Hyperfine structure}
\label{subsection.hfs_theory}
The hyperfine contribution to the Hamiltonian is represented by a multipole
expansion 
\begin{equation}
\label{H_hfs}
  H_{\rm{hfs}}=\sum_{k\geq1} {\bf T}^{(k)} \cdot {\bf M}^{(k)},
\end{equation}
where ${\bf T}^{(k)}$ and ${\bf M}^{(k)}$ are spherical tensor operators
of rank $k$ in the electronic and nuclear spaces, respectively.
The $k=1$ and $k=2$ terms represent the M1 and the E2 interactions. 
In the fully relativistic approach, 
the electronic  contributions are obtained 
from the expectation values of the irreducible spherical
tensor operators~\cite{LindgrenRosen:74a,Jonetal:96c}
\begin{equation}
\label{T1_R}
 \mbox{\bf T} ^{(1)} = -{\rm i} \alpha \sum_{j=1}^{N_e} 
 \left( {\bm{ \alpha }}_j \cdot
  \mbox{\bf l}_j \;  \mbox{\bf C}^{(1)}(j)  \right)  \frac{1}{ r_j^2} \; ,
\end{equation}
and
\begin{equation}
\label{T_2}
 \mbox{\bf T} ^{(2)} =  - \sum_{j=1}^{N_e} \mathbf{C}^{(2)}(j) \frac{1}{r_j^{3}} \; ,
\end{equation}
where $\mbox{\bf l}$ is the electronic orbital angular momentum and $\mbox{\bf C}^{(1)},\mbox{\bf C}^{(2)}$ are the renormalized spherical harmonics of rank $1$ and rank $2$, respectively. 
The EFG, also denoted $q$~\cite{Pyykko:97c}, is obtained from the reduced matrix element of the operator~(\ref{T_2}) using the electronic wave function of the electronic state in question (see Refs.~\cite{Jonetal:96c,Bieron:e-N:2015} for details). It corresponds to the electronic part of the E2 hyperfine interaction constant $B$. The latter is expressed in units of MHz and can be calculated  using the following equation
\begin{equation}
\label{B_frequency_barns_3}
B/\mbox{MHz } =  234.9646 \, (q/a_0^{-3}) (Q/\mbox{b}) \; ,
\end{equation}
where the EFG, or $q$, and the nuclear quadrupole moment, $Q$, are expressed in $a_0^{-3}$ and barns, respectively.
We introduce the isotope-independent hyperfine constants of an electronic state of total angular momentum $J$ as follows
\begin{equation}
\label{A_el} 
A_{\text{el}} \equiv \frac{1}{\sqrt{J(J+1)(2J+1)}} \; \langle \Psi||\mbox{\bf T} ^{(1)}||\Psi\rangle=~ A \left( \frac{I}{\mu_I} \right) ~[\mbox{MHz/}\mu_N] \; ,
\end{equation}
\begin{equation}
\label{B_el} 
B_{\text{el}} \equiv 2\sqrt{\frac{J(2J-1)}{(J+1)(2J+1)(2J+3)}} \;\langle \Psi||\mbox{\bf T} ^{(2)}||\Psi\rangle=~ B/Q ~[\mbox{MHz/b}] \;. \vspace{0.3cm}
\end{equation}
In the equations above, we adopted the definition of the reduced matrix element, which is compatible with the Wigner-Eckart theorem of Edmonds \cite{Edm:57a},
as used in most of the atomic physics textbooks~\cite{GGJ:note}.
The electronic quantities \eqref{A_el} and \eqref{B_el} are closely related to the hyperfine constants $A$ and $B$, which can then easily be computed for a given isotope characterized by the ($\mu_I$, $I$, $Q$) set of nuclear parameters.


\section{S-MR-MCDHF calculations}
\label{section.Heidelberg}
In this first set of calculations, the initial approximation of the atomic states was acquired by using {\sc Grasp2018}~\cite{GRASP2018} to perform an MCDHF calculation on expansions that are built from a set of reference configurations.
For this calculation, we used an equally weighted set of the $5s^2 5p6s$ $^3\!P_1^{\rm o}$ and $5s^2 5p6s$ $^1\!P_1^{\rm o}$ configuration states, together with the lowest $5s^2 5p5d$ $J^{\Pi}=1^-$ state, which is close to the $5s^2 5p6s$ $J^{\Pi}=1^-$ levels. Following~\cite{Dem1984}, mixings with the $5s 5p^3$ configurations were also taken into account, since these were found to strongly influence the odd levels of Sn~I~\cite{Dem1984}. These configurations -- $5s^2 5p6s$, $5s^2 5p5d$, $5s 5p^3$, represented by nine CSFs  --  were treated in the \enquote{extended optimal level} (EOL) scheme~\cite{Grant1980}. As will be discussed below, the aforementioned set of reference configurations was further extended.

The  spectroscopic (occupied) orbitals that took part in the initial MCDHF calculation were kept fixed in all subsequent MCDHF (and RCI) calculations. As a next step, virtual orbitals were generated in MCDHF calculations based on CSF expansions that were produced by allowing single (S) electron substitutions from all spectroscopic orbitals to the subspace of virtual orbitals. Therefore, we label the method presented in this section as S-MR-MCDHF, where S stands for the single electron substitutions and MR indicates that more than one reference configuration were accounted for. 

Due to the one-body nature of the hyperfine operators \eqref{T1_R} and \eqref{T_2}, the S substitutions are known to play an important role in the calculations of hyperfine structures, which is, also, in agreement with the perturbative analysis conducted by, e.g., Verdebout {\it et~al.}~\cite{Verdebout:2013}. CSFs generated by S electron substitutions interact, i.e., the corresponding matrix elements of the hyperfine operators are non-zero, with at least one of the CSFs that are built from the MR configurations. Based on similar arguments, the triple (T) substitutions are quite as crucial. The T substitutions may be decomposed into single and double (SD) substitutions followed by S substitutions. This implies that the CSFs built from configurations that differ by T electron substitutions from the configurations in the MR interact through the one-body hyperfine operators with the energetically important CSFs that are generated by double (D) substitutions. By using more than one reference configuration, the current computational strategy takes into account important D and T electron substitutions from the targeted $5s^25p6s$ configuration.

The sequence of added layers of virtual orbitals were added is given in column~2 of Table~\ref{tab:table1}. In column~3 of the same table, where the label MR3 underlines the number of 
reference configurations, the resulting numbers of CSFs, $N_\text{CSF}$, are displayed for every additional virtual orbital layer $i$. At every step, the previously generated virtual orbitals were kept fixed and, in the subsequent step, only the newly added virtual orbitals were variationally optimized.
Hence, the $6p$ and $4f$ orbitals that make up the first layer of virtual orbitals were obtained by keeping the core orbitals fixed, the $7s$, $7p$, $6d$, and $5f$ orbitals that correspond to the second virtual orbital layer were generated in the next step by also keeping the $6p$ and $4f$ orbitals fixed, and so forth. Overall, eight virtual orbital layers were built, with the last virtual orbital layer $i=8$ corresponding to the $13s13p12d11f10g9h8i$ set of orbitals.

\begin{table}[!h]
\caption{The sequence of the layers of virtual orbitals that were optimized in the S-MR3-MCDHF and S-MR4-MCDHF calculations. The former optimization scheme is based on S electron substitutions from the MR3 set of reference configurations, i.e., \{$5s^25p6s,5s^25p5d,5s5p^3$\}, whereas the latter scheme also includes the $5s^25p7s$ configuration in the so-called MR4 multi-reference. When all four configurations are included in the MR, the $7s$ orbital is part of the spectroscopic orbitals and it is, thus, placed in parentheses in row~3, which displays the $i=2$ virtual orbital layer. In columns~3 and~4, the numbers of generated CSFs, $N_\text{CSF}$, are, respectively, given for each of the two different optimization strategies.}
\label{tab:table1}
\centering
\begin{tabular}{lclcrcr}
\hline
\hline
&& && \multicolumn{3}{c}{$N_\text{{CSF}}$} \\
\cline{5-7}
$i$ && Layers of virtual orbitals                && MR3     && MR4 \\ \hline
\vspace*{0.1cm}
        &&none (MR)  && 9       && 11       \\ 
1       && $6p$, $4f$                                      && 2\,097   && 2\,479     \\
2       && ($7s$,) $7p$, $6d$, $5f$                        && 4\,348   && 4\,820     \\
3       && $8s$, $8p$, $7d$, $6f$, $5g$                    && 7\,054   && 7\,886     \\
4       && $9s$, $9p$, $8d$, $7f$, $6g$                    && 9\,759   && 10\,952    \\
5      && $10s$, $10p$, $9d$, $8f$, $7g$, $6h$            && 12\,563  && 14\,120    \\
6      && $11s$, $11p$, $10d$, $9f$, $8g$, $7h$           && 15\,367  && 17\,288    \\
7      && $12s$, $12p$, $11d$, $10f$, $9g$, $8h$, $7i$    && 18\,242  && 20\,529    \\
8      &~~& $13s$, $13p$, $12d$, $11f$, $10g$, $9h$, $8i$ &~~~~& 21\,117 &~~~&23\,770 \\
\hline
\hline
\end{tabular}
\end{table}

\begin{figure*}[!h]
\centerline{\includegraphics[width=0.80\textwidth]{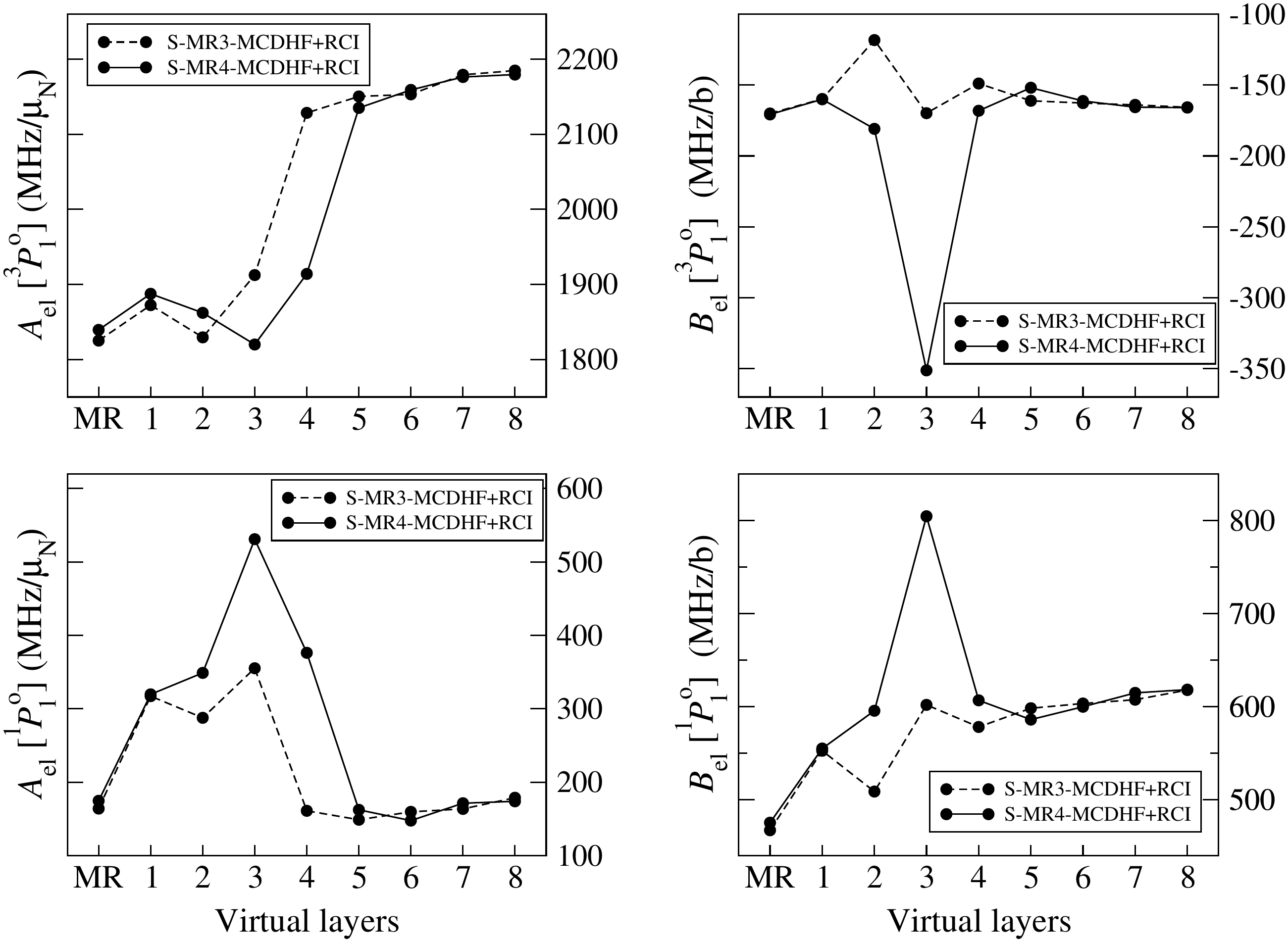}}
\caption{The convergence patterns of the electronic hyperfine factors $A_{\text{el}}[^{3}\!P_{1}^{\rm o}]$, $A_{\text{el}}[^{1}\!P_{1}^{\rm o}]$ (left panels) and $B_{\text{el}}[^{3}\!P_{1}^{\rm o}]$, $B_{\text{el}}[^{1}\!P_{1}^{\rm o}]$ (right panels) as functions of the virtual orbital layers. The radial orbital basis was obtained by applying two different
optimization strategies with respect to the selected MR configurations. The dashed lines connect the values resulting from the S-MR3-MCDHF optimization, where three reference configurations are included in the MR, and the solid lines link the resulting values from the S-MR4-MCDHF scheme, where the MR was extended to include four reference configurations. Both sets of values are the results from the RCI calculations that followed the orbital optimization step. For further details, see text in Sec.~\ref{section.Heidelberg}. 
}
\label{fig:CIconv} 
\end{figure*}

Following the MCDHF calculations, valence-valence (VV) correlations were ultimately included in the RCI calculations by also allowing D substitutions of electrons from a smaller set of valence subshells, i.e., $5s$, $5p$, $5d$, and $6s$, to the space of virtual orbitals. D substitutions from lower-lying subshells were not included to keep the number of CSFs at a manageable level. The resulting values of the isotope-independent hyperfine constants $A_{\text{el}}$ and $B_{\text{el}}$  are shown in Fig.~\ref{fig:CIconv} with the label S-MR3-MCDHF+RCI.
\vspace{0.5cm}

In a succeeding set of calculations, the number of reference configurations was extended to include the $5s^2 5p 7s$ $J^{\Pi}=1^-$ configuration in the initial multi-reference optimization of the spectroscopic orbitals. The sequence of the virtual orbitals that were progressively optimized is similar to the previous calculations (see Table~\ref{tab:table1}). Since, at this point, the $7s$ orbital is part of the spectroscopic orbitals, the second virtual orbital layer consists of the $7p$, $6d$, and $5f$ orbitals alone. For every additional virtual orbital layer, the resulting numbers of CSFs, $N_\text{CSF}$, are displayed in column~4 of Table~\ref{tab:table1}, under the name MR4.  
Likewise the S-MR3-MCDHF calculations, these MCDHF calculations were completed with RCI calculations that included SD electron substitutions from the valence subshells, i.e., $5s$, $5p$, $5d$, $6s$, and $7s$. The respective results of the electronic hyperfine factors are illustrated in Fig.~\ref{fig:CIconv} with the label S-MR4-MCDHF+RCI. 


As seen in Fig.~\ref{fig:CIconv}, the computed electronic hyperfine factors for the two targeted $5s^{2}5p6s \ {}^{1,3}\!P_{1}^{\rm o}$ states are effectively converged. For the largest CSF expansions, no noticeable change is observed between the results from the S-MR3-MCDHF and S-MR4-MCDHF optimization strategies.
That being so, the final results are taken from the largest RCI calculation using the $13s13p12d11f10g9h8i$ orbital basis set that was optimized within the S-MR4-MCDHF scheme (corresponding to 103~403 CSFs). For this first set of calculations, the final values of the isotope-independent hyperfine constants are:
\begin{equation}
\label{eq:zoltan_res}
\begin{split}
A_{\rm el}\left[{}^3\!P^{\rm o}_1\right]&= 2\ 180~{\rm MHz/}\mu_{\rm N}\,;  \quad B_{\rm el}\left[{}^3\!P^{\rm o}_1\right]=-166~{\rm MHz/b}\,;  \\
A_{\rm el}\left[{}^1\!P^{\rm o}_1\right]&= 174~{\rm MHz/}\mu_{\rm N}\ ;   \quad B_{\rm el}\left[{}^1\!P^{\rm o}_1\right]= 622~{\rm MHz/b}\, .
\end{split}
\end{equation}
It is noteworthy that the above-displayed final $A_{\rm el}\left[{}^3\!P^{\rm o}_1\right]$ and $B_{\rm el}\left[{}^1\!P^{\rm o}_1\right]$ values are significantly larger than the results of the initial MR3(MR4) calculations that were based only on the 9(11) reference CSFs. The relative discrepancies amount to $19\%$($19\%$) and $33\%$($31\%$), respectively. The final $A_{\rm el}\left[{}^1\!P^{\rm o}_1\right]$ and $B_{\rm el}\left[{}^3\!P^{\rm o}_1\right]$ values are, however, surprisingly close to the MR3(MR4) results. 


\section{S\texorpdfstring{\MakeLowercase{r}}{r}D-SR-MCDHF calculations}
\label{section.SRcalculations}
\subsection{Convergence of the electronic hyperfine structure factors}
\label{subsection.SRconvergence}
Experimental data (and state compositions) indicate that 
the hyperfine constant $A$ is large for the $^3\! P^{\rm o}_{1}$ state and small for the $^1\! P^{\rm o}_{1}$ state. Oppositely, the EFG is small, in absolute value, for the $^3\! P^{\rm o}_{1}$ state and large for the $^1\! P^{\rm o}_{1}$ state. When spectroscopic data are used for the extraction of nuclear parameters, only these large values are of interest.
Therefore, the single-reference (SR) calculations of this section focus on the \enquote{large} results, presenting only the electronic hyperfine factors $A_{\text{el}}$[$^3\! P^{\rm o}_{1}$] and $B_{\text{el}}$[$^1\! P^{\rm o}_{1}$]~$\propto$~EFG[$^1\! P^{\rm o}_{1}$] in the following Fig.~\ref{figure.fig2} and Table~\ref{SRtable.a3p1}. 

All calculations were performed together for the targeted $5s^2 5p 6s \,\, ^3\! P^{\rm o}_{1}$
and $5s^2 5p 6s \,\, ^1\! P^{\rm o}_{1}$ excited states.
The spectroscopic and virtual orbitals were optimized based on an EOL energy functional that was built over the CSFs of the two lowest $J^{\Pi}=1^{-}$ levels. 
The generation of the wave functions essentially followed the scheme described in our previous papers~\cite{Bieron1999BeF,BieronAu2009,Bieron:zinc4s4p:2018}.
The virtual orbitals were generated in a
series of SCF calculations, where the virtual orbital set was systematically increased by one layer. Each layer included one additional virtual orbital of the
$s$, $p$, $d$, $f$, and $g$ angular symmetries.
In the very last, eighth layer, only the $s$, $p$, $d$, and $f$ symmetries were represented. Therefore, the largest multiconfiguration expansion was built on the $14s13p12d11f11g$
set of orbitals.  

The occupied orbital shells were systematically
opened for substitutions, starting from the $5p$ and $6s$ valence orbitals and, eventually, opening all occupied orbital shells, down to the $1s$ core orbital. All virtual orbitals were generated in calculations with single and restricted double substitutions (SrD), 
which means (1) unrestricted single substitutions
and (2) double substitutions restricted by the limitation
that at most one electron might be substituted from occupied, closed shells
with  $n<5$.
The other (or both) electron(s) must be substituted from the  $5s$, $5p$, and $6s$ subshells.
The largest expansion 
contains substitutions from all occupied orbital shells  
to eight layers of virtual orbitals of the $s$, $p$, $d$, $f$, and $g$ symmetries, except for the very last layer that was limited to $l_{max}=3$ as mentioned above. 

The orbitals generated in this first phase were frozen and used in subsequent RCI calculations that constituted the next two phases of the calculations.
The second phase allowed single and (unrestricted) double substitutions (SD), while the third phase 
allowed single, double, and triple substitutions (SDT). For the latter two phases, only six layers of virtual orbitals were retained,  since both $A_{\text{el}}$[$^3\! P^{\rm o}_{1}$] and $B_{\text{el}}$[$^1\! P^{\rm o}_{1}$] values saturated at the 7$^{th}$ and 8$^{th}$ layers of the first phase, as also illustrated in Fig.~\ref{figure.fig2} (see magenta circles). 
Accordingly, the $A_{\text{el}}$[$^3\! P^{\rm o}_{1}$] and $B_{\text{el}}$[$^1\! P^{\rm o}_{1}$] values resulting from the 6$^{th}$ layer of phase~1 are taken as a reference and shown in the second row of Table~\ref{SRtable.a3p1}. In the same table, the first row displays the $A_{\text{el}}$[$^3\! P^{\rm o}_{1}$] and $B_{\text{el}}$[$^1\! P^{\rm o}_{1}$] values from the DHF computation, restricted to two CSFs. The DHF values differ substantially from the converged values of phase 1, i.e., by $26\%$ and $25\%$, for $A_{\text{el}}$[$^3\! P^{\rm o}_{1}$] and $B_{\text{el}}$[$^1\! P^{\rm o}_{1}$], respectively. The large deviations from the simplest DHF calculations validate the significance of electron correlation effects in the computation of the electronic hyperfine structure factors.

\begin{figure*}
\begin{minipage}{0.48\textwidth}
\includegraphics[width=1\textwidth]{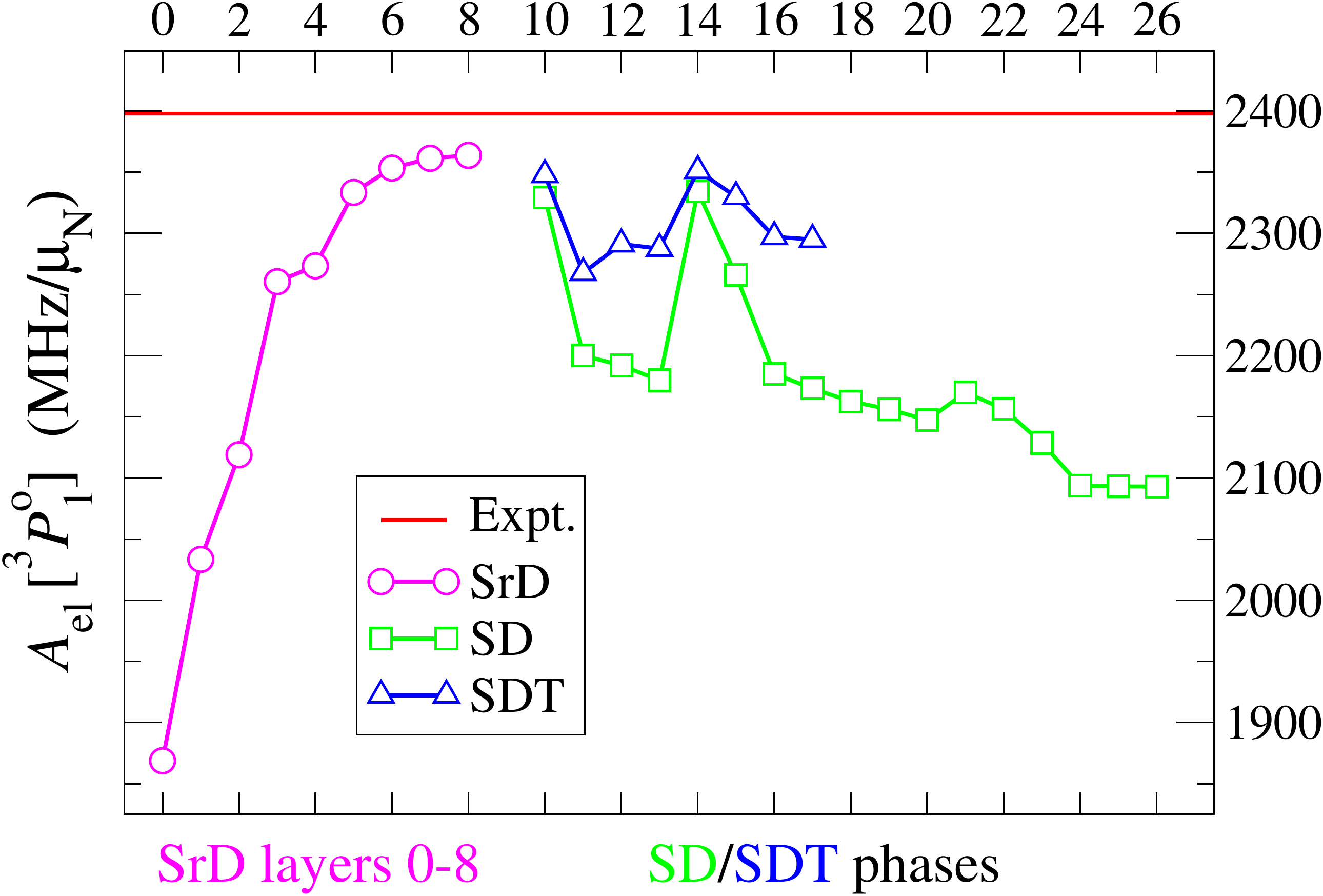}
\end{minipage}\hfill \hfill \hfill
\begin{minipage}{0.47\textwidth}
\includegraphics[width=1\textwidth]{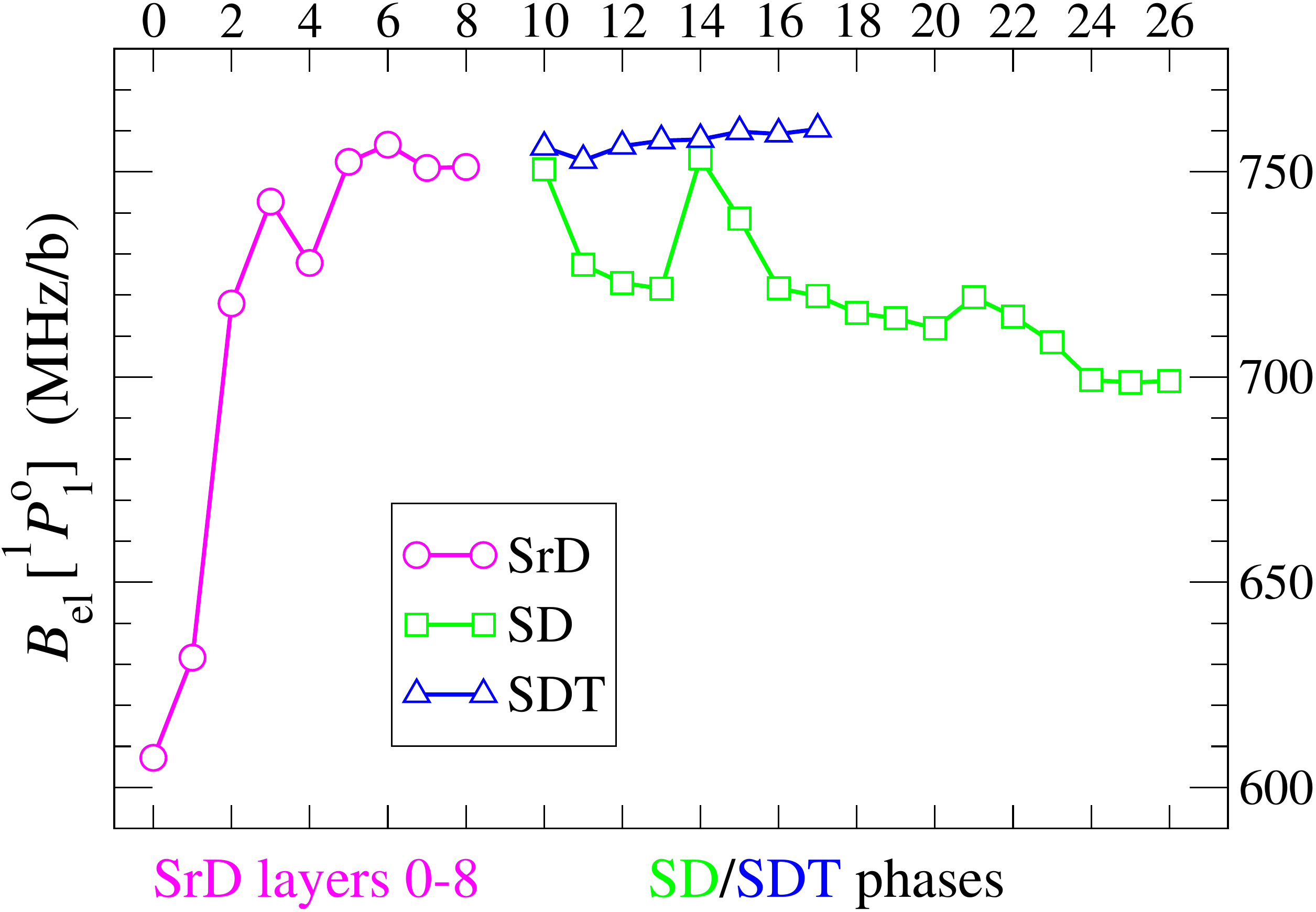}
\end{minipage}
\vspace{0.35cm}
\caption{(Color online) The electronic hyperfine factors $A_{\text{el}}$[$^3\! P^{\rm o}_{1}$]~(in MHz/$\mu_N$) and $B_{\text{el}}$[$^1\! P^{\rm o}_{1}$]~(in MHz/b) obtained in three approximations: the SrD (magenta circles), the SD (green squares), and the
SDT (blue triangles) computational approaches. 
The symbols represent the results at each step of the calculations, while the lines are only for the guidance of the eyes. On the x-axes, the number $0$ indicates the DHF computation, the numbers 1-8 represent the consecutive layers of virtual orbitals
developed in the SrD phase of the calculations, and the numbers 10-26 match the labels of the multiconfiguration expansions presented in the
first column of Table~\ref{SRtable.a3p1},
corresponding to the calculations performed in the SD and SDT phases.
The red straight horizontal line on the top of the left graph represents the experimental value
$A_{\text{el}}^{\text{expt}}[{}^3\!P^{\rm o}_{1}]$~=~2\ 398~MHz/$\mu_N$ from Ref.~\cite{Yordanov:20}. For further details, see text in Sec.~\ref{section.SRcalculations}.}
\label{figure.fig2}
\end{figure*}

Similarly to any other expectation value, the isotope-independent hyperfine constants $A_{\text{el}}$[$^3\! P^{\rm o}_{1}$] and $B_{\text{el}}$[$^1\! P^{\rm o}_{1}$] are functions of four variables, called \enquote{dimensions} in Ref.~\cite{BieronAu2009}. These \enquote{dimensions} refer to
the number of virtual orbital layers, the virtual space angular momentum, the opened occupied shells, and the number of substitutions, i.e., S, D, T, quadruple (Q), or higher.
In the second, or else SD, phase, the multiconfiguration expansions
were systematically increased in the first three of the \enquote{dimensions} above.
The resulting values of the electronic hyperfine factors $A_{\text{el}}$[$^3\! P^{\rm o}_{1}$] and $B_{\text{el}}$[$^1\! P^{\rm o}_{1}$] from the SD phase are presented in Table~\ref{SRtable.a3p1} as functions of the increasing multiconfiguration expansions. The computed values are also displayed graphically in Fig.~\ref{figure.fig2}, in which the points on the $x$ axis represent the labels displayed in the first column of Table~\ref{SRtable.a3p1}. For each labelled calculation, the second column of Table~\ref{SRtable.a3p1} gives the principal quantum number $n$ of the deepest orbital shell that was opened for substitutions. For instance, $n\geq4$ involves substitutions from the $4s,4p,4d,5s,5p$, and $6s$ orbitals. The third column of the same table provides the active set of orbitals to which the electron substitutions were allowed.

\begin{table*} 
\caption{
The computed electronic hyperfine factors $A_{\text{el}}[{}^3\!P_1^{\rm o}]$ (in MHz/$\mu_N$) and $B_{\text{el}}[{}^1\!P_1^{\rm o}]$ (in MHz/b) for various multiconfiguration expansions. The considered CSFs were generated based on SD (columns 4 and 6) and SDT (columns 5 and 7) substitutions from the opened shells displayed in column~2 to the active set of orbitals given in column 3. The first row contains the resulting $A_{\text{el}}[{}^3\!P_1^{\rm o}]$ and $B_{\text{el}}[{}^1\!P_1^{\rm o}]$ values from the DHF computation, where only the CSFs of the two targeted states were considered, and the second row displays the converged results from the SrD computational phase after the sixth layer of virtual orbitals was added. The labels given in column~1 correspond to the labels used on the horizontal axes of Fig.~\ref{figure.fig2}.
}
\label{SRtable.a3p1}
\begin{tabular}{clcrllccccccc}
\hline
\hline
& & & & & & \multicolumn{3}{c}{$A_{\text{el}}[{}^3\!P_1^{\rm o}]$ (MHz/$\mu_N$)}&&
\multicolumn{3}{c}{$B_{\text{el}}[{}^1\!P_1^{\rm o}]$ (MHz/b)} \\
\cline{7-9}
\cline{11-13}
Label &&
 \multicolumn{1}{c}{open shells} &
 \multicolumn{1}{c}{{\phantom{empty}}} &
 \multicolumn{1}{l}{active orbital set} &
 \multicolumn{1}{c}{{\phantom{empty}}} &
 \multicolumn{1}{c}{SD} & & \multicolumn{1}{c}{SDT}&&\multicolumn{1}{c}{SD} &&
 \multicolumn{1}{c}{SDT}\\
  \hline
\vspace*{0.1cm} 
0 & \multicolumn{4}{c}{phase 1: DHF computation~~} && \multicolumn{3}{c}{\!\!\!\!\!\!\!\!\!\!1\,869} &&
 \multicolumn{3}{c}{\!\!\!\!\!\!\!\!\!\!607} \\
\vspace*{0.1cm} 
6 & \multicolumn{4}{c}{phase 1: SrD virtual layer 6} && \multicolumn{3}{c}{\!\!\!\!\!\!\!\!\!\!2\,353} && \multicolumn{3}{c}{\!\!\!\!\!\!\!\!\!\!757} \\ 
10 && $n\geq5$   && $8s7p$      && 2\,329 && 2\,348 && 751 && 756\\
11 && $n\geq4$ && $7s6p5d4f$    && 2\,200 && 2\,268 && 727 && 753\\
12 && $n\geq4$ && $8s7p5d4f$  && 2\,192 && 2\,291 && 723 && 756\\
13 && $n\geq4$ && $8s7p5d4f5g$  && 2\,180 && 2\,288 && 722 && 758\\
14 && $n\geq4$ && $9s8p$      && 2\,335 && 2\,351 && 753 && 758\\
15 && $n\geq4$ && $9s8p5d$   && 2\,266 && 2\,330 && 739 && 760\\
16 && $n\geq4$ && $9s8p5d4f$   && 2\,185 && 2\,297 && 722 && 759\\
17 && $n\geq4$ && $9s8p5d4f5g$  && 2\,173 && 2\,295 && 720 && 760\\
18 && $n\geq4$ && $10s9p6d5f6g$  && 2\,163 && /       && 716 && / \\
19 && $n\geq4$ && $11s10p7d6f7g$  && 2\,156 && /      && 714 && / \\
20 && $n\geq4$ && $12s11p10d9f10g$   && 2\,147 && /   && 712 && / \\
21 && $n\geq3$ && $9s8p5d4f5g$  && 2\,170 && / && 719 && / \\
22 && $n\geq3$ && $10s9p6d5f6g$  && 2\,157 && / && 715 && / \\
23 && $n\geq3$ && $11s10p7d6f7g$  && 2\,129 && / && 709 && / \\
24 && $n\geq3$ && $12s11p10d9f10g$   && 2\,094 && / && 699 && / \\
25 && $n\geq3$ && $13s12p11d10f11g$   && 2\,093 && / && 699 && / \\
26 &~~~& $n\geq2$ && $12s11p10d9f10g$   && 2\,089 &~~~& / &~~~~~~~~~& 698 &~~~& / \\
&& Expt.~\cite{Yordanov:20}    &&&& \multicolumn{3}{c}{\!\!\!\!\!\!2\,398(2)} &&&\\
\hline
\hline
\end{tabular}
\end{table*}

%
Saturation of the SD phase is observed for the $n\geq3 \rightarrow 12s11p10d9f10g$
calculation (see label~24 in Table~\ref{SRtable.a3p1}).
The saturation is demonstrated by comparing the results obtained
from the label-24 calculation
with the results obtained from the
two larger expansions:
the label-25 calculation,  which was extended by one additional virtual orbital layer, and the label-26 calculation, where substitutions from the  $n=2$ shells were added. 
All these three different expansions yield similar values and,
therefore, the expansion with label~24 ($n~\geq3 \rightarrow 12s11p10d9f10g$)
was carried over to the SDT phase of the calculations.

In the third phase, involving SDT electron substitutions, we tried to follow a pattern for generating the multiconfiguration expansions that was similar to the SD phase. However, the number of the generated CSFs based on T substitutions was growing very rapidly and the limits of the computational resources available to us were reached before the computed properties were fully saturated. The results from the SDT phase are presented alongside the SD results in Table~\ref{SRtable.a3p1} and graphically in Fig.~\ref{figure.fig2}.
It should be mentioned that  the SDT calculations included the CSFs produced during both the 
SrD phase ($n\geq1 \rightarrow 12s12p10d9f10g$) and the SD phase ($n\geq3 \rightarrow 12s11p10d9f10g$).

\subsection{Correlation between $A_{\text{el}}$[$^3\! P^{\rm o}_{1}$] and $B_{\text{el}}$[$^1\! P^{\rm o}_{1}$]}
\label{section.SRresults}

The electronic M1 hyperfine factor $A_{\text{el}}$ appears to be quite sensitive to D and T substitutions. 
The dependence of the computed $A_{\text{el}}$ values on the different classes of electron substitutions, i.e., S, D, and T substitutions, is illustrated in the left graph of Fig.~\ref{figure.fig2}.
The overall dependence follows the trends that had been observed in many earlier
calculations~\cite{Engels1993,Engels1996,PJMRG,MRGPJ,Bieron:Au:2008,BieronAu2009,Bieron:zinc4s4p:2018},
where the effect of the D substitutions was to decrease the absolute values of the computed electronic hyperfine factors and the effect of the T substitutions was to decrease the effect of the D substitutions.
In fact, these dependencies are visible in Fig.~\ref{figure.fig2} for both $A_{\text{el}}$[$^3\! P^{\rm o}_{1}$] and $B_{\text{el}}$[$^1\! P^{\rm o}_{1}$]. 
%

The comparison between the two graphs of Fig.~\ref{figure.fig2} further
illustrates the correlation between the
$A_{\text{el}}$[$^3\! P^{\rm o}_{1}$] and $B_{\text{el}}$[$^1\! P^{\rm o}_{1}$], which is evident not only in the SrD phase of the calculations (magenta circles), but also in the case of the SD approximation (green squares). 
This observation will be exploited in Sec.~\ref{section.Estimates} as a tool for determining a more reliable value for the electronic E2 hyperfine factor $B_{\text{el}}$[$^1\! P^{\rm o}_{1}$].  As seen in Fig.~\ref{figure.fig2}, the correlation between $A_{\text{el}}$[$^3\! P^{\rm o}_{1}$] and $B_{\text{el}}$[$^1\! P^{\rm o}_{1}$] is less pronounced in the case of SDT approximation (blue triangles). It appears that $B_{\text{el}}$[$^1\! P^{\rm o}_{1}$] is insensitive to T substitutions. However, the comparison between the results obtained in the
SD and SDT approximations shows that the T substitutions are still quite effective in decreasing the effects of the D substitutions.

\subsection{Estimating $B_{\text{el}}$[$^1\! P^{\rm o}_{1}$]}
\label{section.Estimates}

It is often the case that an estimate of an error bar of a calculated
expectation value of a particular atomic property~$X$ is based on a calculated value of another atomic property~$Y$.
This situation often occurs when the accuracy of $X$ cannot be easily obtained, while the accuracy of $Y$ can be obtained from comparison with experiment or by other means.
Examples of such indirect estimations include: error bars of transition rates~\cite{Ekmetal:2014a,Ekmetal:2014b} or isotope shifts~\cite{Bisetal:2016a,Filetal:2017a} inferred from the accuracy of the corresponding transition energies; 
error bars in the calculations of amplitudes involved in parity- and time-reversal symmetry violations inferred from hyperfine calculations~\cite{GingesFlambaum2004,RadziuteRaHgYbEDM2014};
error bars of E2 hyperfine constants inferred from calculations of M1 hyperfine constants~\cite{BieronAu2009,Bieron:zinc4s4p:2018}.

In the case of the hyperfine interaction constants, 
when a series of multiconfiguration expansions are employed,
both $A_{\text{el}}$ and $B_{\text{el}}$ factors exhibit similar and synchronous dependence on the size of the expansion.
This is illustrated in Fig.~\ref{figure.fig2} and in numerous previous calculations of hyperfine 
structures~\cite{BieronLi1996,BieronBeF1999,BieronBi2001,BieronHg2005,Froemmgen:Cd:2015}. Synchronous oscillations that are observed in the computed $A_{\text{el}}$ and $B_{\text{el}}$ values may be qualitatively explained by comparing two consecutive multiconfiguration expansions. Within the MCDHF methodology employed in the present paper, two neighboring multiconfiguration expansions,
$N_{\text{CSF}}({i})$ and $N_{\text{CSF}}({i\!+\!1})$, 
differ by the presence of an additional $i\!+\!1$ layer of virtual orbitals 
in the $N_{\text{CSF}}({i+1})$ expansion.
%
The orbitals represented in the $N_{\text{CSF}}({i})$ expansion are frozen,
and only the orbitals included in the additional $i\!+\!1$ layer 
are optimized in the SCF procedure described in Sec.~\ref{section.Theory}.
The additional layer of virtual orbitals introduces contributions to the expectation values of the M1 and E2 hyperfine operators, and these contributions, in turn, depend  on the radial forms of the additional virtual orbitals. 
A \enquote{layer} is composed of a set of virtual orbitals
with principal quantum numbers increased by one with respect
to the previous layer. 
Each layer includes orbitals with different angular quantum numbers
(typically one orbital per angular symmetry is represented in each consecutive layer). 
The evaluation of hyperfine structures involves radial integrals, in which
a radial factor $r^{-3}$ induces a strong dependence on the inner part
of the electronic orbitals~\cite{Kopfermann1958,LindgrenRosen:74a,FBJbook}, which results in the
well known statement that the hyperfine interaction happens within 
the innermost one-half of the first oscillation of one-electron 
orbitals~\cite{Pyykko1972,Pyykko1973}.
This common $r^{-3}$ dependence of the radial hyperfine integrals for both M1 and E2 interactions,
although not immediately obvious from Eqs.~(\ref{T1_R}) and (\ref{T_2}),
may be explained from the
different structures of the relevant one-electron matrix elements~\cite{JohnsonBook:2007}.
The additional layer of virtual orbitals
induces a $\Delta A$ shift of the $A$ constant and, respectively, a $\Delta B$  shift of the $B$ constant (for the purpose of this discussion we used the M1 hyperfine constant $A$
and the E2 hyperfine constant $B$,
but the arguments 
apply similarly to
the isotope-independent hyperfine constants $A_{\text{el}}$ and $B_{\text{el}}$).
From the above considerations one should expect that
these shifts are approximately proportional, i.e.,
\begin{equation}
\label{dAdB}
 \Delta A / A \approx \Delta B / B .
\end{equation}
This may be illustrated in Fig.~8 of Ref.~\cite{Verdebout:2013},
where the curves $a_{\text{dip}}$ (top right) and $b_{\text{quad}}$ (bottom right)
apparently oscillate synchronously as functions of the
multiconfiguration expansion, and in numerous previous calculations of hyperfine 
structures~\cite{BieronLi1996,BieronBeF1999,BieronBi2001,BieronHg2005,Froemmgen:Cd:2015}.

The above equation may be transformed into a relation in which the calculated values $A_{\text{calc}}$ and $B_{\text{calc}}$ are related to the experimental values $A_{\text{expt}}$ and $B_{\text{expt}}$, so that
%
\begin{equation}
\label{AexptBcalc}
 |A_{\rm calc}-A_{\text{expt}}| / A_{\text{expt}} \approx |B_{\text{calc}}-B_{\text{expt}}| / B_{\text{expt}}.
\end{equation}
The equation above can, then, be used to correct the calculated 
value of the E2 hyperfine factor $B_{\text{el}}$
(or the calculated EFG~$\propto B_{\text{el}}$)
by a semiempirical shift arising from the (known) error in the
calculated value of the M1 hyperfine factor $A_{\text{el}}$.

As already discussed in Sec.~\ref{subsection.SRconvergence}, the resulting $A_{\text{el}}$[$^3\! P^{\rm o}_{1}$] and $B_{\text{el}}$[$^1\! P^{\rm o}_{1}$] values from the SD approximation were fully converged, whereas the multiconfiguration calculations involving SDT substitutions were not entirely saturated due to the exceptionally large CSF expansions. 
It should be pointed out that the largest completed SDT calculation, with 4~406~086 CSFs, took 37 days of wall time on the computer cluster at our disposal (6x96~CPU @~2.4GHz with 6x256~GB~RAM). The next in line, with 17~817~617 CSFs, eventually exceeded the capacity of that cluster.
%
However, as mentioned in Sec.~\ref{section.SRresults} and illustrated in Fig.~\ref{figure.fig2}, $B_{\text{el}}$[$^1\! P^{\rm o}_{1}$] is rather insensitive to T substitutions. On this ground, the $B_{\text{el}}$[$^1\! P^{\rm o}_{1}] = 760$~MHz/b value obtained from the largest completed SDT calculation is still taken into consideration in the analysis below for determining the final $B_{\text{el}}$[$^1\! P^{\rm o}_{1}$] value of the SrD-SR-MCDHF and RCI calculations described in this section. In the following, the latter result is labeled $B_{\text{el}}\text{(SDT)}$. 

When applying the semiempirical shift of Eq.~\eqref{AexptBcalc} to the final results obtained from the SrD, SD, and SDT calculations described in Sec.~\ref{section.SRcalculations}, three more $B_{\text{el}}$[$^1\! P^{\rm o}_{1}$] values are obtained, which are labeled $B_{\text{el}}(\text{SrD})^{\rm shifted}$, $B_{\text{el}}(\text{SD})^{\rm shifted}$, and $B_{\text{el}}(\text{SDT})^{\rm shifted}$, respectively. To estimate these three values, the experimental $A_{\text{el}}^{\rm expt}[^{3}\!P_{1}^{\rm o}]$~=~2\ 398~MHz/$\mu_N$ value was taken from Ref.~\cite{Yordanov:20}.
We ultimately arrive at the following four figures:
$B_{\text{el}}(\text{SDT})$~=~760~MHz/b,
$B_{\text{el}}(\text{SrD})^{\rm shifted}$~=~759~MHz/b,
$B_{\text{el}}(\text{SD})^{\rm shifted}$~=~800~MHz/b and
$B_{\text{el}}(\text{SDT})^{\rm shifted}$~=~793~MHz/b.
%
By taking their average, we obtain:
\begin{equation}\label{eq:result778}
B_{\text{el}}[^1\! P^{\rm o}_{1}]~=~778~{\text{MHz/b}}.    
\end{equation}

\section{S\texorpdfstring{\MakeLowercase{r}}{r}D-MR-MCDHF calculations}
\label{section.MRcalculations}

\subsection{Selecting the multi-reference set}
\label{section.MRset}
In this third approach, we go beyond the single-reference approximation described in Sec.~\ref{section.SRcalculations} by defining a set of MR configurations. These configurations must be selected so that they produce a number of CSFs that account for the major electron correlation effects, or else static correlation~\cite{CFFreview:2016}. Further, by allowing S and D substitutions of electrons from the MR configurations, important T and Q substitutions from the targeted $5s^{2}5p6s$ configuration are taken into account. Starting from single-reference SD-MCDHF calculations, where the two targeted $5s^{2}5p6s \ {}^{1,3}\!P_{1}^{\rm o}$ states were optimized according to the EOL scheme, the $LS$-composition of the resulting ASFs was analyzed. After allowing, for instance, SD substitutions from the valence orbitals ($n\geq5$) to a first layer of virtual orbitals, i.e., $7s,6p,6d$, and $4f$, the $LS$-composition of each of the targeted states can be seen in Table~\ref{tab:ls_composition}. From the first analysis based on Table~\ref{tab:ls_composition}, we, thus, defined an MR that is composed of the following three non-relativistic configurations: $5s^{2}5p6s$, $5p^{3}6s$, and $5s5p5d6s$.
The above configurations correspond to 17 relativistic CSFs for the $J^{\Pi} = 1^{-}$ symmetry. Enlarging further the MR set would lead to very large CSF expansions that are beyond the reach of the computational resources that are available to us. For that reason, the MR was restricted to the three leading configurations.

\begin{table*}
\caption{The $LS$-composition of the two targeted $5s^{2}5p6s \ {}^{1,3}\!P_{1}^{\rm o}$ states after performing initial SD-MCDHF calculations (see also text in Sec.~\ref{section.MRset}). The percentages of the four most dominant $LS$-components are solely displayed, with the first percentage corresponding to the assigned configuration and term.}
\label{tab:ls_composition}      
\centering          
\begin{tabular}{c c c c}
\hline
\hline
Pos. & Conf. & ~$LSJ$ & $LS$-composition \\
\hline
1    & ~$5s^{2}5p6s$ & ~$^3\!P_1^{\rm o}$ & ~~~0.761 + 0.186~$5s^{2}5p6s~^1\!P^{\rm o}$ + 0.016~$5p^36s~^3\!P^{\rm o}$ + 0.010~$5s5p5d6s~^3\!P^{\rm o}$  \\
2    & ~$5s^{2}5p6s$ & ~$^1\!P_1^{\rm o}$ & ~~~0.761 + 0.185~$5s^{2}5p6s~^3\!P^{\rm o}$ + 0.017~$5p^36s~^1\!P^{\rm o}$ + 0.009~$5s5p5d6s~^1\!P^{\rm o}$  \\
\hline
\hline
\end{tabular}
\end{table*}

\subsection{Orbital optimization strategies} 
\label{section.orb_opt}

The active set of virtual orbitals was systematically increased by one layer, with each layer~$i$ including orbitals with quantum numbers $nl$ equal to $(i+6)s,(i+5)p,(i+5)d,$ and $(i+3)f$.
Therefore, the last virtual orbital layer $i=9$ corresponds to the $15s14p14d12f$ set of orbitals. 
The orbital basis was obtained by performing two different sets of MCDHF computations, which are, respectively, denoted VV-SD-MR-MCDHF and CV-SrD-MR-MCDHF. The former orbital optimization strategy included SD substitutions from the $5s,5p,5d,$ and $6s$ valence orbitals of the MR configurations, capturing valence-valenve (VV) correlation effects. On the other side, the latter optimization strategy allowed SrD substitutions from the $4d,5s,5p,5d,$ and $6s$ orbitals in the MR, with the limitation that there was at most one substitution from the $4d$ subshell. All other inner subshells were kept closed. Since the $4d$ orbital is considered part of the core, the generated CSF expansions based on the above-mentioned SrD substitutions further accounted for core-valence (CV) correlation effects.

\begin{figure*}
\includegraphics[width=0.8\textwidth]{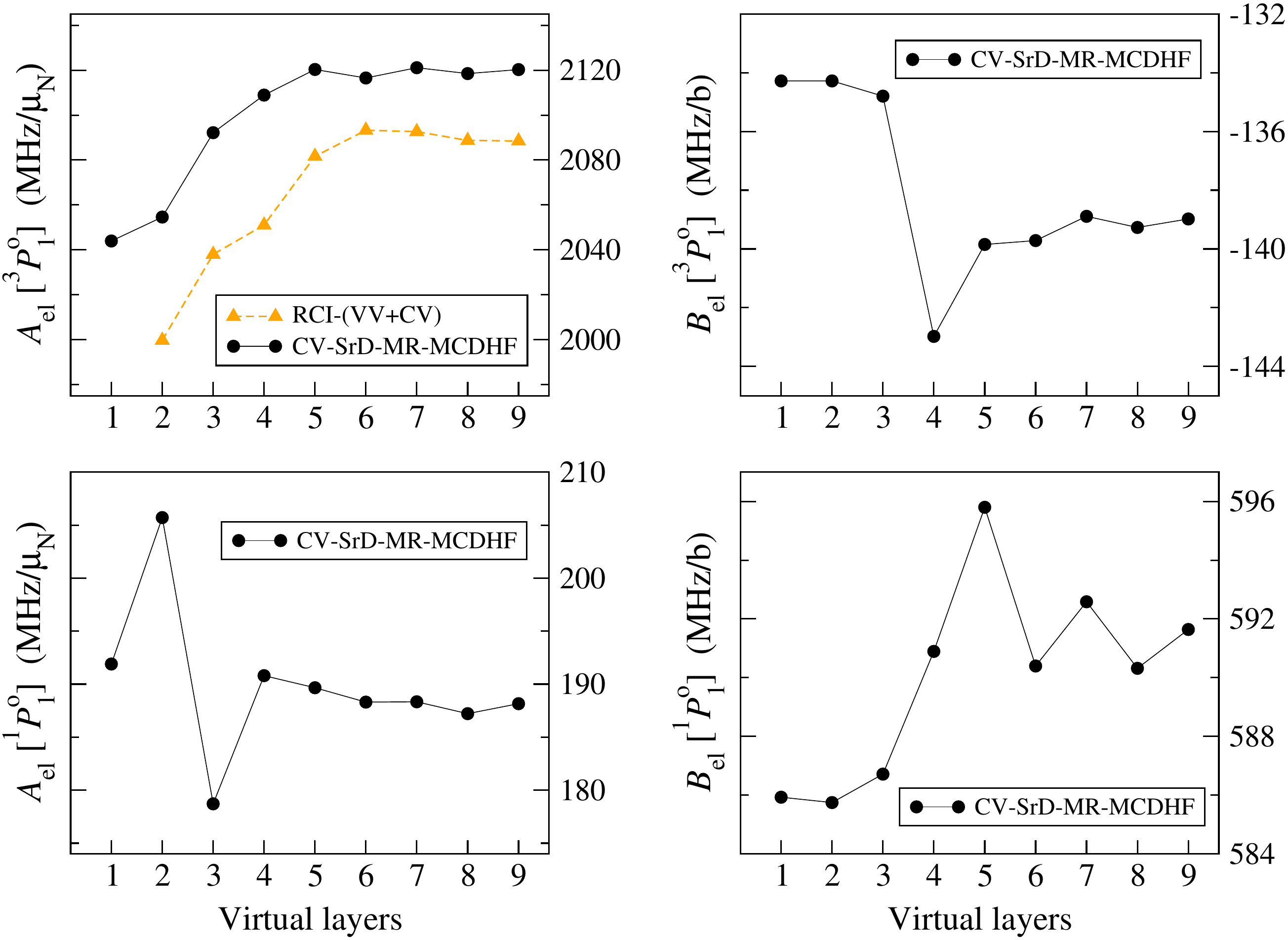}
\caption[]{(Color online) The convergence of the electronic hyperfine factors $A_{\text{el}}[^{3}\!P_{1}^{\rm o}]$, $A_{\text{el}}[^{1}\!P_{1}^{\rm o}]$ (left panels) and $B_{\text{el}}[^{3}\!P_{1}^{\rm o}]$, $B_{\text{el}}[^{1}\!P_{1}^{\rm o}]$ (right panels) with the increasing number of virtual orbital layers, optimized using the CV-SrD-MR-MCDHF scheme. In the upper left panel, the computed $A_{\text{el}}[^{3}\!P_{1}^{\rm o}]$ values using the CV-SrD-MR-MCDHF scheme (black circles) are compared with the $A_{\text{el}}[^{3}\!P_{1}^{\rm o}]$ values resulting from the VV-SD-MR-MCDHF orbital optimization strategy and additional RCI-(VV+CV) computations, where CV correlation was included (orange triangles). For details, see text in Secs.~\ref{section.orb_opt} and~\ref{section.convergence}.}
\label{fig:hfs_MR}
\end{figure*}

By using the orbitals optimized in the VV-SD-MR-MCDHF scheme and, for each virtual orbital layer, performing an RCI calculation based on the CSF expansion that was generated in the CV-SrD-MR-MCDHF computations, the values of the electronic factor
$A_{\text{el}}$[${}^{3}\!P_{1}^{\rm o}$] were compared for the two different orbital optimization strategies. The upper left panel of Fig.~\ref{fig:hfs_MR} illustrates the convergence of $A_{\text{el}}$[${}^{3}\!P_{1}^{\rm o}$] as a function of the number of virtual orbital layers for both sets of computations; the results from the VV-SD-MR-MCDHF optimization strategy are represented by RCI-(VV+CV). Although the two approaches exhibit similar trends, the discrepancies between the resulting $A_{\text{el}}[{}^{3}\!P_{1}^{\rm o}]$ values strongly suggest to use the CV-SrD-MR-MCDHF orbital set to perform additional RCI computations (see Sec.~\ref{section.final_RCI}). After adding nine layers of virtual orbitals, the absolute discrepancy is $\sim32$~MHz/$\mu_{\text{N}}$. As will be seen, this value is much larger than the variations induced by the effects of the \enquote{$f$-limit} and the reduction of the CSF space, which are, respectively, investigated in the following two subsections.

\subsection{The \enquote{$f$-limit}}
\label{section.flimit}

The angular quantum number $l$ of the orbital basis was limited to $l_{max}=3$ so that only orbitals up to the $f$ symmetry were included in the computations presented in this section. 
The effects of orbitals with higher angular symmetry on the computation of the electronic hyperfine factors are, generally, known to be small~\cite{SunOls:93a}. In the present work, this was verified by comparing two sets of VV-SD-MR-MCDHF computations that, respectively, utilized $l_{max}=3$ and $l_{max}=5$ (the latter includes orbitals of the $g$ and $h$ symmetries). The VV-SD-MR-MCDHF computational scheme results in smaller number of CSFs (compared to the CV-SrD-MR-MCDHF scheme) and it, therefore, serves better the purpose of this subsection, 
which is to quantify the error $\epsilon_f$ induced by the choice of the \enquote{$f$-limit}. 

These VV-SD-MR-MCDHF computations were carried out for the first five virtual orbital layers, and the resulting $A^{h}_{\text{el}}[{} ^3\!P^{\rm o}_{1}]$ and $A^f_{\text{el}}[{} ^3\!P^{\rm o}_{1}]$ values, respectively, corresponding to the $l_{max}=5$ and $l_{max}=3$ active spaces, are presented in Table~\ref{tab:flim}. Looking at Table~\ref{tab:flim}, after adding five layers of virtual orbitals, the difference between the two sets of values is $\Delta A_{\text{el}}^f[{} ^3\!P^{\rm o}_{1}]=$ $A_{\text{el}}^h$[${} ^3\!P^{\rm o}_{1}] - A_{\text{el}}^f$[${} ^3\!P^{\rm o}_{1}]\simeq3$ MHz/$\mu_{\text{N}}$. A similar analysis yields $\Delta A_{\text{el}}^f[{} ^1\!P^{\rm o}_{1}]\simeq1$~MHz/$\mu_{\text{N}}$, $\Delta B_{\text{el}}^f[{}^{3}\!P^{\rm o}_{1}]\simeq2$~MHz/b, and $\Delta B_{\text{el}}^f[{}^{1}\!P^{\rm o}_{1}]\simeq0$~MHz/b. 
The differences $\Delta A_{\text{el}}^f$ and $\Delta B_{\text{el}}^f$ are expected to remain almost unvarying as the number of virtual orbital layers increases further, and when presenting the final results of the computed electronic hyperfine factors in Sec.~\ref{section.final_RCI}, the above-mentioned values are assumed to be the corrections $\epsilon_f$ due to the \enquote{$f$-limit}.

\begin{table}[H]
\centering
\caption[]{The influence of orbitals with $g$ and $h$ angular symmetries on the computation of the electronic hyperfine factor $A_{\text{el}}[{}^3\!P^{\rm o}_{1}]$. 
By, respectively, using $l_{max}=5$ and $l_{max}=3$, the $A^{h}_{\text{el}}[{}^3\!P^{\rm o}_{1}]$ and $A^f_{\text{el}}[{}^3\!P^{\rm o}_{1}]$ values were computed in MHz/$\mu_{\text{N}}$ for five virtual orbital layers that were optimized in the VV-SD-MR-MCDHF scheme, and the results are, respectively, shown in columns~2 and 3. In the last column, the differences $\Delta A_{\text{el}}^f[{} ^3\!P^{\rm o}_{1}]=A_{\text{el}}^h$[${} ^3\!P^{\rm o}_{1}]-A_{\text{el}}^f$[${} ^3\!P^{\rm o}_{1}]$ are also presented.}
\label{tab:flim}
\begin{tabular}{cccc}
\hline
\hline
Virtual layer & ~$A_{\text{el}}^h[{}^3\!P^{\rm o}_{1}]$& ~$A_{\text{el}}^f[{}^3\!P^{\rm o}_{1}]$ & ~$\Delta A_{\text{el}}^f[{} ^3\!P^{\rm o}_{1}]$\\
\hline
1 & 2\,046  &  2\,044  &  2 \\
2 & 2\,054  &  2\,052  &  2 \\
3 & 2\,074  &  2\,071  &  3 \\
4 & 2\,060  &  2\,058  &  ~3$^{\dagger}$ \\
5 & 2\,064  &  2\,061  &  3 \\
\hline
\hline
\multicolumn{4}{l}{\footnotesize{$^\dagger$Note that the roundings were performed}}\\
\vspace*{-1cm} \\
\multicolumn{4}{l}{\footnotesize{after evaluating the difference.}}
\end{tabular}
\end{table}

\subsection{Reduction of CSF space}
\label{section.RED}


When we open the $4d$ subshell to include CV correlation in the CV-SrD-MR-MCDHF optimization strategy, the number of CSFs grows very rapidly with the increasing active set of virtual orbitals. It is, then, advisable to restrict the atomic state expansions to CSFs that interact, i.e., have non-zero Hamiltonian matrix elements, with the ones generated by the MR configurations. This was done by utilizing the tool \textsf{rcsfinteract}, which is included in both {\sc Grasp2K} \citep{grasp2K:2013} and {\sc Grasp2018} \citep{GRASP2018} computer packages. Such reduction of CSFs normally does not bring any major losses in accuracy~\cite{Caretal:2010a,Lietal:2012c,Gaietal:2020a}.

\begin{table*}
\centering
\caption[]{The effect of reducing the configuration space, to CSFs that interact with the ones that are part of the MR, on the computation of the electronic hyperfine factor $A_{\text{el}}[{}^3\!P^{\rm o}_{1}]$. Using the full and reduced CSF spaces, two different sets of computations were performed, in which nine virtual orbital layers were optimized in the VV-SD-MR-MCDHF scheme. The numbers of CSFs before and after the reduction are, respectively, given in columns~2 and~3. The corresponding $A^{\rm full}_{\text{el}}[{}^3\!P^{\rm o}_{1}]$ and $A_{\text{el}}^{\text{red}}[{}^{3}\!P^{\rm o}_{1}]$ values are given in MHz/$\mu_{\text{N}}$ in columns~4 and~5, and their discrepancies $\Delta A_{\text{el}}^{\text{red}}[{} ^3\!P^{\rm o}_{1}]$ are displayed in the last column. In the second portion of the table, results from additional RCI-(VV+CV) computations up to the fifth virtual orbital layer are included.}
\label{fig:interact}
\begin{tabular}{ccccccccccc}
\hline
\hline
Virtual layer & N$_\text{CSF}$ &&N$^{\text{red}}_\text{CSF}$&& $A^{\rm full}_{\text{el}}$[${}^3\!P^{\rm o}_{1}$] && $A^{\text{red}}_{\text{el}}$[${}^3\!P^{\rm o}_{1}$] & ~$\Delta A_{\text{el}}^{\text{red}}[{} ^3\!P^{\rm o}_{1}]$  \\
\hline
2&~4\,069&&~3\,192&& 2\,052  &&  2\,045  &  ~6$^\dagger$ \\
3&~8\,400&&~6\,518&& 2\,071  &&  2\,063  &  8 \\
4&14\,293&&11\,008&& 2\,058  &&  2\,051  &  7  \\
5&21\,748&&16\,662&& 2\,061  &&  2\,053  &  8 \\
6&30\,765&&23\,480&&2\,053  &&  2\,046  &  7\\
7&41\,344&&31\,462&&2\,054  &&  2\,047  &  ~8$^\dagger$\\
8&53\,485&&40\,608&&2\,049  &&  2\,041  &  ~7$^\dagger$\\
9&67\,188&&50\,918&&2\,049  &&  2\,041  &  ~7$^\dagger$ \\
\hline
RCI-(VV+CV) &&& \\
2&~71\,747&&~57\,086&& 2\,007 & &  2\,000  &  ~8$^\dagger$  \\
3&149\,060&&122\,610&& 2\,047 & &  2\,038  &  9 \\
4&254\,485&&212\,946&& 2\,059 & &  2\,051  &  8  \\
5&388\,022&~~~&328\,094&~~~& 2\,089 & &  2\,082  &  7 \\
\hline
\hline
\multicolumn{9}{l}{\footnotesize{$^\dagger$Note that the roundings were performed after evaluating the difference.}}
\end{tabular}
\end{table*}

The effect of limiting the number of CSFs to the \enquote{interacting} ones is evaluated by comparing the results from two sets of computations, where for each one the full and reduced CSF spaces were used. Similarly to the investigation of the effect of the \enquote{$f$-limit}, the electronic hyperfine factors were computed in the VV-SD-MR-MCDHF scheme, which is computationally faster. The resulting $A^{\rm full}_{\text{el}}$[${}^{3}\!P^{\rm o}_{1}$] and $A^{\rm red}_{\text{el}}$[${}^{3}\!P^{\rm o}_{1}$] values, respectively, corresponding to the full and reduced CSF spaces, together with the associated numbers of generated CSFs, are displayed in Table~\ref{fig:interact}. Looking at Table~\ref{fig:interact}, we observe that, after building nine layers of virtual orbitals, the discrepancy $\Delta A_{\text{el}}^{\text{red}}[{}^{3}\!P^{\rm o}_{1}]=A^{\rm full}_{\text{el}}[{}^{3}\!P^{\rm o}_{1}]-A_{\text{el}}^{\text{red}}[{}^{3}\!P^{\rm o}_{1}]$ converges to about $7$~MHz/$\mu_{\text{N}}$. A similar analysis yields $\Delta A_{\text{el}}^{\text{red}}[{}^{1}\!P^{\rm o}_{1}] \simeq -5$~MHz/$\mu_{\text{N}}$, $\Delta B_{\text{el}}^{\text{red}}[{}^{3}\!P^{\rm o}_{1}] \simeq -1$~MHz/b, and $\Delta B_{\text{el}}^{\text{red}}[{}^{1}\!P^{\rm o}_{1}] \simeq 0$~MHz/b.

To further evaluate the possible influence that the reduction of the CSF space might have on CV correlation, additional RCI-(VV+CV) computations were performed including five virtual orbital layers. The resulting $A^{\rm full}_{\text{el}}$[${}^{3}\!P^{\rm o}_{1}$] and $A_{\text{el}}^{\text{red}}[{}^{3}\!P^{\rm o}_{1}]$ values from the latter RCI computations, together with the numbers of the considered CSFs, are displayed in the second portion of Table~\ref{fig:interact}. It is seen that $\Delta A_{\text{el}}^{\text{red}}[{}^{3}\!P^{\rm o}_{1}]$ is kept almost unchanged and, thus, unaffected by CV correlation, and the same applies to $\Delta A_{\text{el}}^{\text{red}}[{}^{1}\!P^{\rm o}_{1}]$. That being so, when inferring the final values of the computed electronic hyperfine factors $A_{\text{el}}$ in Sec.~\ref{section.final_RCI}, it is assumed that the corrections due to the reduction of the CSF space, $\epsilon_\text{red}$, are, respectively, equivalent to $\Delta A_{\text{el}}^{\text{red}}[{}^{3}\!P^{\rm o}_{1}]=7$~MHz/$\mu_{\text{N}}$ and $\Delta A_{\text{el}}^{\text{red}}[{}^{1}\!P^{\rm o}_{1}]=-5$~MHz/$\mu_{\text{N}}$, just as estimated above. On the other hand, a noticeable increase in the variations of the electronic factors $B_\text{el}$ was observed and, for those, it was concluded that the corrections $\epsilon_\text{red}$ equal $\Delta B_{\text{el}}^{\text{red}}[{}^{3}\!P^{\rm o}_{1}]=-1$~MHz/b and $\Delta B_{\text{el}}^{\text{red}}[{}^{1}\!P^{\rm o}_{1}]=1$~MHz/b, respectively.

\subsection{Convergence of the CV-SrD-MR-MCDHF computations}
\label{section.convergence}

The importance of optimizing the orbital basis using the CV-SrD-MR-MCDHF computational strategy, which also takes into account the polarization of the $4d$ core orbital through CV substitutions, was already highlighted in Sec.~\ref{section.orb_opt} (see also Fig.~\ref{fig:hfs_MR}). The purpose of this subsection is to examine the convergence of the properties of interest within the CV-SrD-MR-MCDHF approach. In Table~\ref{table:hfs_MR_conv}, the computed excitation energies and electronic hyperfine factors $A_{\text{el}}$ and $B_{\text{el}}$ of the $^{1,3}\!P_{1}^{\rm o}$ states are given as functions of the increasing active set of virtual orbitals. As a reference point, the values associated with the initial MR calculation are also given in the first row of Table~\ref{table:hfs_MR_conv}. One notices that these values are close to the resulting $A_{\text{el}}[{}^{3}\!P^{\rm o}_{1}]$ and $B_{\text{el}}[{}^{1}\!P^{\rm o}_{1}]$ values from the DHF calculations presented in Table~\ref{SRtable.a3p1}. 
Table~\ref{table:hfs_MR_conv} shows that nine virtual orbital layers are required to ensure the convergence of all computed properties. We note that, in the final atomic state expansion for the $J=1^{-}$ symmetry, the number of CSFs exceeds one million.

\begin{table}[h!]
\caption{The convergence of energies and electronic hyperfine factors $A_{\text{el}}$ and $B_{\text{el}}$ for the $^{1,3}\!P_{1}^{\rm o}$ states after extending the MR orbital basis to include nine layers of virtual orbitals that were optimized using the CV-SrD-MR-MCDHF scheme. The computed excitation energies of the $^{3}\!P_{1}^{\rm o}$ and $^{1}\!P_{1}^{\rm o}$ states are, respectively, presented in columns~2 and 3, whereas the evaluated energy separations are displayed in column~4. For comparison, the observed energies are shown at the bottom part of the table. In each of the columns~5 and~6, the values of the electronic factors $A_{\text{el}}[^{3}\!P_{1}^{\rm o}]$ and $A_{\text{el}}[^{1}\!P_{1}^{\rm o}]$ are given, and columns~7 and~8, respectively, contain the values of the electronic factors $B_{\text{el}}[^{3}\!P_{1}^{\rm o}]$ and $B_{\text{el}}[^{1}\!P_{1}^{\rm o}]$. The last column exhibits the numbers of generated CSFs for every additional virtual orbital layer.}
\label{table:hfs_MR_conv}      
\centering    
\begin{tabular}{ccccccccccccccr}
\hline
\hline
& \multicolumn{4}{c}{Energies (cm$^{-1}$)} && \multicolumn{3}{c}{$A_{\text{el}}$ (MHz/$\mu_\text{N}$)} &&	\multicolumn{3}{c}{$B_{\text{el}}$ (MHz/b)} && \\
\cline{2-5}
\cline{7-9}
\cline{11-13}
Virtual layer~~ & $^3\!P_{1}^{\rm o}$ && $^1\!P_{1}^{\rm o}$ &$^1\!P_{1}^{\rm o}-{}^3\!P_{1}^{\rm o}$ && $^3\!P_{1}^{\rm o}$&& $^1\!P_{1}^{\rm o}$&& $^3\!P_{1}^{\rm o}$&& $^1\!P_{1}^{\rm o}$ && $N_{\text{CSFs}}$ \\
\hline
\vspace*{0.1cm}
none (MR)  &  33\,301 &&  38\,002 & 4\,701 && 1\,869  && 515  &&  $-$154 && 613 && 17 \\
1  &  34\,327 &&  38\,990 & 4\,663 && 2\,044  && 192  &&  $-$134 && 586 && 16\,593  \\
2  &  34\,789 &&  39\,323 & 4\,534 && 2\,055  && 206  &&  $-$134 && 586 && 57\,086  \\
3  &  34\,644 &&  39\,154 & 4\,510 && 2\,092  && 179  &&  $-$135 && 587 && 122\,610  \\
4  &  34\,587 &&  39\,072 & 4\,485 && 2\,109  && 191  &&  $-$143 && 591 && 212\,946  \\
5  &  34\,544 &&  39\,020 & 4\,476 && 2\,120  && 190  &&  $-$140 && 596 && 328\,094  \\
6  &  34\,529 &&  39\,004 & 4\,475 && 2\,117  && 188  &&  $-$140 && 590 && 468\,054  \\
7  &  34\,521 &&  38\,993 & 4\,472 && 2\,121  && 188  &&  $-$139 && 593 && 632\,826  \\
8  &  34\,519 &&  38\,987 & 4\,468 && 2\,119  && 187  &&  $-$139 && 590 && 822\,410  \\
9  &  34\,517 &~~&  38\,984 & 4\,467 &~~& 2\,120 &~~& 186 &~~~&  $-$139 &~~& 592 &~~& 1\,036\,806\\
&&&& \\  
Expt.~\cite{NIST_ASD,Brill1964}&  34\,914 &&  39\,257 & 4\,343 & \\
\hline
\hline
\end{tabular}
\end{table}

At this point, the excitation energies of the $^{3}\!P_{1}^{\rm o}$ and $^{1}\!P_{1}^{\rm o}$ states are well converged and the predicted energy separation between the two states agrees with the observed value to within $3\%$ (see column~4 in Table~\ref{table:hfs_MR_conv}). The quality of the obtained energies provides an initial assessment of the computed electronic hyperfine factors, which, as seen in Table~\ref{table:hfs_MR_conv}, are also effectively converged. The convergence patterns of the $A_{\text{el}}[{}^{3}\!P^{\rm o}_{1}]$, $A_{\text{el}}[{}^{1}\!P^{\rm o}_{1}]$, $B_{\text{el}}[{}^{3}\!P^{\rm o}_{1}]$, and $B_{\text{el}}[{}^{1}\!P^{\rm o}_{1}]$ values with respect to the increasing number of virtual orbital layers are also illustrated in Fig.~\ref{fig:hfs_MR}. One should bear in mind that the y-axes in the two right-hand side graphs of Fig.~\ref{fig:hfs_MR} span a considerably smaller range of values compared to the graphs on the left side. That being so, once the sixth virtual layer has been added, the convergences of the $A_{\text{el}}$ and $B_{\text{el}}$ factors should be considered equally smooth.

\subsection{Final CV-SrD-MR-MCDHF $+$ RCI computations}
\label{section.final_RCI}

After ensuring the convergence of the computed properties within the CV-SrD-MR-MCDHF orbital optimization scheme, a final RCI computation was carried out, 
in which the atomic state expansions were augmented to include CSFs accounting for additional electron correlation effects. Due to limited computational power, the CSF space must be generated so that the most dominant correlation effects are, first and foremost, efficiently captured. To assess the relative importance of the various correlation effects on the computation of the electronic hyperfine factors, preliminary RCI computations, which used the $10s9p9d7f$ set of orbitals optimized within the CV-SrD-MR-MCDHF scheme, were performed, and their results are discussed below.

\begin{table}
\caption{The effect of different types of electron substitutions on the computation of the electronic hyperfine factors $A_{\text{el}}$ and $B_{\text{el}}$ for the $^{1,3}\!P_{1}^{\rm o}$ states. 
The reference values displayed in the first line were computed after optimizing four virtual orbital layers using the CV-SrD-MR-MCDHF scheme (maximum one hole allowed in the $4d$ orbital).
Contributions from additional substitutions were evaluated in subsequent RCI computations, in which the configuration space was enlarged by, first, including~D substitutions from the $4d$ orbital and, then, progressively adding S substitutions from the $4p$ orbital down to the $1s$ orbital.
In the last portion of the table, contributions from CV correlation effects were evaluated in RCI computations, where the configuration space was generated by allowing SrD substitutions from $n\geq4$ orbitals with the restriction that there was at most one substitution, initially, only from the $4p$ orbital, and then, from the $4s$ orbital as well.
}
\label{table:core_exc}      
\centering          
\begin{tabular}{lccccccccccc}
\hline
\hline		
& & \multicolumn{3}{c}{$A_{\text{el}}$ (MHz/$\mu_{\text{N}}$)} && 	\multicolumn{3}{c}{$B_{\text{el}}$ (MHz/b)}  \\
\cline{3-5}
\cline{7-9}
Subst. & ~~Orbital~~ & $^3\!P_{1}^{\rm o}$&& $^1\!P_{1}^{\rm o}$&& $^3\!P_{1}^{\rm o}$ && $^1\!P_{1}^{\rm o}$\\
\hline \midrule
SrD & $4d$    &  2\,109 &&  191 && $-$143 && 591 \\ \hline \midrule
SD & $4d$    &  1\,996 &&  308 &&  $-$149 && 605 \\
$+$ S  & $4p$     &  2\,058  &&  332 && $-$166 && 671 \\
$+$ S  & $4s$     &  2\,062  &&  379 && $-$166 && 671\\
$+$ S  & $3d$    &  2\,061  &&  379 &&  $-$167 && 672 \\
$+$ S  & $3p$ &  2\,081  &&  387 &&  $-$170 && 684 \\
$+$ S  & $3s$ &  2\,086  &&  389 &&  $-$170 && 684 \\
$+$ S  & $2p$ &  2\,094  &&  392 &&  $-$170 && 687 \\
$+$ S  & $2s$ &  2\,097  &&  389 &&  $-$170 && 687 \\
$+$ S  & $1s$ &  2\,099  &&  388 &&  $-$170 && 687 \\[.4cm]
 \hline \midrule
SD & $4d$    & 1\,996 &&  308 &&  $-$149 && 605 \\
$+$ SrD & $4p$   &  2\,087 &&  334 &&  $-$164 && 674 \\
$+$ SrD & $4s$ &  2\,095 &~~~&  378 &~~~~~&  $-$164 &~~~& 675\\
\hline
\hline
\end{tabular}
\end{table}

VV and some CV electron correlation effects were already captured during the optimization of the orbital basis. To also account for core (C) and core-core (CC) correlation effects, inner subshells were progressively opened to allow substitutions of electrons to the $10s9p9d7f$ active set of orbitals. At first, additional D substitutions were allowed from the $4d$ subshell. Then, S substitutions from inner core subshells were gradually considered, starting from the $4p$ subshell and eventually opening all occupied subshells down to the innermost $1s$ core orbital. Further CV correlation effects were considered by allowing rD substitutions from the $n\geq4$ orbitals with the limitation that there is maximum one substitution, initially, only from the $4p$ orbital, and then, also from the $4s$ orbital. For every additional type of electron substitutions described above, RCI computations using the respective CSF expansions were performed. For each of these RCI computations, the resulting values of the electronic hyperfine factors $A_{\text{el}}$ and $B_{\text{el}}$ are shown in Table~\ref{table:core_exc}.

Looking at Table~\ref{table:core_exc}, the D substitutions from the $4d$ orbital are clearly the most important to consider for the final RCI computation. The S substitutions from $np$ orbitals also have a rather substantial effect, although it becomes less significant as $n$ decreases. All S substitutions from $ns$ orbitals are less important, but computationally cheap, and they were, thus, considered in the CSF expansion of the final RCI computation. Looking at the last portion of Table~\ref{table:core_exc}, one realizes that the CV correlation effects from both $4p$ and $4s$ subshells are important. However, our computational resources allowed us to include rD substitutions only from the $4p$ orbital, and not from the $4s$ orbital, when constructing the final expansions of the atomic states. 

The final RCI computation was carried out by applying the above-mentioned rules of electron substitutions to the largest active set, corresponding to the $15s14p14d12f$ set of orbitals. 
The number of CSFs in the final expansions was $3~583~001$. In Table~\ref{table:final}, the concluding values of the electronic hyperfine factors $A_{\text{el}}$ and $B_{\text{el}}$ are presented together with the corrections $\epsilon_{\rm red}$ and $\epsilon_{f}$, respectively, induced by the reduction of the configuration space (see Sec.~\ref{section.RED}) and by the omission of $l\geq4$ orbitals in the active set (see Sec.~\ref{section.flimit}).

\begin{table}[H]
\caption{The resulting values of the electronic hyperfine factors $A_{\text{el}}$ and $B_{\text{el}}$ for the $^3\!P_{1}^{\rm o}$ and $^1\!P_{1}^{\rm o}$ states from the final CV-SrD-MR-MCDHF+RCI computations. The corrections $\epsilon_{red}$ and $\epsilon_{f}$ are further added to account for reducing the active space to configurations that interact with the MR and for limiting the angular quantum number $l$ of the orbitals basis to $l_{max}=3$ 
.}
\label{table:final}      
\centering    
\begin{tabular}{lccccccc}
\hline
\hline
&&\multicolumn{2}{c}{$A_{\text{el}}$ (MHz/$\mu_{\text{N}}$)} & &	\multicolumn{2}{c}{$B_{\text{el}}$ (MHz/b)}  \\
\cline{3-4}
\cline{6-7}
&&$^3\!P_{1}^{\rm o}$& $^1\!P_{1}^{\rm o}$&~~~& $^3\!P_{1}^{\rm o}$& $^1\!P_{1}^{\rm o}$ \\
\hline
Final RCI &&2\,169 &  409 && $-$173 &  716 \\
\quad$+\, \epsilon_{\rm red}$ &&2\,176 & 404 && $-$174 & 717 \\
\quad$+ \, \epsilon_{f}$ &~~~&2\,179 & 405 && $-$172 & 717 \\
\hline
\hline
\end{tabular}
\end{table}

\subsection{Sensitivity to orbital bases and CSF expansions}
\label{section.sensitivity}
As seen in the previous sections, different calculations based on the same general method, the MCDHF method, and performed with the same program, the {\sc Grasp2018} package, lead to different results. The $A_{\text{el}}$[$^3\!P_{1}^{\rm o}$], $B_{\text{el}}$[$^3\!P_{1}^{\rm o}$], and $B_{\text{el}}$[$^1\!P_{1}^{\rm o}$] values are in agreement within a relative error of approximately $10\%$, while the $A_{\text{el}}$[$^1\!P_{1}^{\rm o}$] values differ significantly. The differences in the S-MR-MCDHF, SrD-SR-MCDHF, and SrD-MR-MCDHF approaches lie in the choice of their respective orbital bases and CSF expansions, each with its benefits and drawbacks. In this subsection, we investigate the sensitivity of the SrD-SR-MCDHF and SrD-MR-MCDHF approaches by arbitrarily interchanging their orbital bases and CSF expansions.

\begin{table*}
\centering    
\caption{The electronic hyperfine factors $A_{\rm el}$ and $B_{\rm el}$ for the $^3\!P_{1}^{\rm o}$ and $^1\!P_{1}^{\rm o}$ states, computed for six different combinations of orbital basis sets and CSF spaces. The SrD-SR-MCDHF and SrD-MR-MCDHF computational approaches are compared by expanding the total wave function over the largest CSF expansion of the one method and using the orbital basis of the other method. Adjustments were made in the CSF expansions due to the specific properties of the orbital bases obtained in the two different approaches. For details, see text in Sec.~\ref{section.sensitivity}.}
\label{table:sensitivity} 
\begin{tabular}{llccccccccc}
\hline
\hline		
 & &\multicolumn{2}{c}{$A_{\text{el}}$ (MHz/$\mu_{\text{N}}$)} && 	\multicolumn{2}{c}{$B_{\text{el}}$ (MHz/b)}  
 & \\
 \cline{3-4}
 \cline{6-7}
 Orb. basis~~ & CSF expansions & $^3\!P_{1}^{\rm o}$& $^1\!P_{1}^{\rm o}$&& $^3\!P_{1}^{\rm o}$ & $^1\!P_{1}^{\rm o}$
 & ~~$B_{\text{el}}$[$^3\!P_{1}^{\rm o}]/B_{\text{el}}[^1\!P_{1}^{\rm o}$] \\
\hline
SrD-MR & SrD-MR & 2\,179 & 404 && $-$172 & 717 & $-$0.240\\
SrD-SR & SrD-SR & 2\,295 & 285 && $-$190 & 760 &$-$0.250\\[.3cm]
SrD-MR & SrD-SR($spdf$ limit)~ & 2\,303&289 &&$-$188 & 739&$-$0.254 \\
SrD-SR & SrD-SR($spdf$ limit)~ & 2\,297 &  289 && $-$190 & 722&$-$0.263 \\[.3cm]
SrD-MR & SrD-MR($6$ layers) & 2\,161 & 396 && $-$172 & 709&$-$0.243 \\
SrD-SR & SrD-MR($6$ layers) & 2\,168 & 342 &~~~& $-$194 & 718&$-$0.270\\
\hline
\hline
\end{tabular}
\end{table*}

Each of the above-mentioned methods provides the final results by ultimately performing RCI computations. As explained before, the largest RCI expansion in the CV-SrD-MR-MCDHF method accounts for electron substitutions to the $15s14p14d12f$ set of orbitals including S substitutions from all occupied orbitals, D substitutions from the $4d$, $5s$, and $5p$ subshells, and rD substitutions from the $4p$ subshell. The SrD-SR-MCDHF approach allows all D and T substitutions from orbital shells with $n=4,5$, and $6$ to three layers of $s$ and $p$ virtual orbitals and to one layer of $d,f$, and $g$ virtual orbitals. 
This was merged with the CSFs generated by allowing SrD substitutions from all core subshells to six layers of $s,p,d,f,$ and $g$ virtual orbitals. 
These results are presented in Table~\ref{table:sensitivity} under the labels SrD-MR/SrD-MR and SrD-SR/SrD-SR, for the SrD-MR-MCDHF and SrD-SR-MCDHF approaches, respectively (where the notation X/Y defines the orbital basis from X and CSF expansion from Y). 
Two additional sets of computations were performed: one combining the SrD-SR-MCDHF orbital basis and the SrD-MR-MCDHF CSF space (see SrD-SR/SrD-MR in Table~\ref{table:sensitivity}) and one combining the SrD-MR-MCDHF orbital basis and the SrD-SR-MCDHF CSF space (see SrD-MR/SR-SrD in Table~\ref{table:sensitivity}). Minor changes in the CSF spaces were required, e.g.,  the SR active space was restricted to the $s,p,d$, and $f$ symmetries as the MR orbital basis is limited to $l_{max}=3$ and the SrD-MR-MCDHF CSF space was limited to only six virtual layers. 

Table~\ref{table:sensitivity} shows that the results are consistent, although far to be in perfect agreement. The effect of replacing the orbital set for a given electron correlation model (i.e., for a given CSF expansion) is (surprisingly) small, which is reassuring. The largest discrepancies  between the two models (SrD-SR-MCDHF and SrD-MR-MCDHF) are observed for $A_{\text{el}}$[$^1\!P_{1}^{\rm o}$] (40\%) and $B_{\text{el}}$[$^3\!P_{1}^{\rm o}$] (10\%). The extreme sensitivity of $A_{\text{el}}$[$^1\!P_{1}^{\rm o}$] is expected, as shown in Appendix~\ref{section.MCDHFBP} utilizing the framework of non-relativistic theory. A similar analysis for the $B_{\text{el}}$[$^3\!P_{1}^{\rm o}]/B_{\text{el}}[^1\!P_{1}^{\rm o}$] ratio reveals a steadier value, which can be derived from a simple $^{3}\!P^{\rm o} - {}^{1}\!P^{\rm o}$ mixing. This analysis is consistent with the $B_{\text{el}}$[$^3\!P_{1}^{\rm o}]/B_{\text{el}}[^1\!P_{1}^{\rm o}$] ratios presented in Table~\ref{table:sensitivity} and the computed $B_{\text{el}}$[$^3\!P_{1}^{\rm o}]/B_{\text{el}}[^1\!P_{1}^{\rm o}$] ratio from the final $B_{\text{el}}$ values given by Eq.~\eqref{eq:zoltan_res} in Sec.~\ref{section.Heidelberg}. The values of all these ratios range from $-0.240$ to $-0.270$, which is in excellent agreement with the experimental value: $-0.25(2)$~\cite{Yordanov:20}.

\section{CI-DFS calculations}
\label{section.CIDFS_calculations}
In this last set of calculations, which is based on the CI-DFS theory, we used for all Sturmian functions the same reference energy, namely, that of the $5s$ state. All orbitals up to $3d$ form the occupied shells' space RAS1, orbitals with $n=4,5$ and the $6s, 6p$ orbitals belong to the active space RAS2, and orbitals from $6d$ and beyond form the open shells' space RAS3. To assess the uncertainty of the calculations presented in this section, we performed a series of calculations using an increasing orbital basis set. SD substitutions from the occupied shells' space and from the open shells' space, i.e., $N_{\rm h}=N_{\rm ex}=N_{\rm el}=2$ (see Sec.~\ref{subsubsubsection.RAS} for their definition), were included in the calculations, leading to a large number of configurations and huge matrices for the numerical diagonalization.
By freezing the $1s$, $2s$, and $2p$ states, and by using PT to build low-lying closed shells and highly-excited states, we were able to extend the one-electron basis to the $12s11p10d9f$ set of orbitals. 
For the three smallest orbital basis sets, T substitutions into RAS2, i.e., $N_{\rm ex}=3$, were also included, however, their influence turned out to be smaller than the uncertainty level we aim at.

\begin{table}
\begin{center}
\caption{The numbers of configurations and the resulting energy separations between the targeted $^3\!P_1^{\rm o}$ and $^1\!P_1^{\rm o}$ states for each virtual orbital layer used in the CI-DFS calculations. Two approaches, the direct (full basis) and the one based on perturbation theory (PT), were implemented. $N_{\rm NonRel}$ stands for the numbers of non-relativistic configurations, and when followed by \enquote{(PT)}, the numbers of configurations built using PT are also displayed in parentheses, $N_{\rm Total}$ indicates the total numbers of relativistic configurations, and the energy separations $\Delta E = E[{}^1\!P_1^{\rm o}]-E[{}^3\!P_1^{\rm o}]$ are given in cm$^{-1}$. The numbers of virtual orbital layers that are given in the first column correspond to the labels used on the horizontal axes of Fig.~\ref{Fig:NO_results}, and the respective orbital basis sets are displayed in column~2.  See also text in Secs.~\ref{subsection.CIDFS} and \ref{section.CIDFS_calculations}.} 
\label{Tab:NO_configurations}
\begin{tabular}{ c l r r c p{0.4cm} r r r c}
\hline \hline
Virtual\,\,\,  & Orbital  & $N_{\rm NonRel}$ & ~$N_{\rm Total}$ & ~$\Delta E$ (cm${}^{-1}$) &  & $N_{\rm NonRel}$ & (PT) & ~$N_{\rm Total}$ & ~$\Delta E$ (cm${}^{-1}$)  \\ 
\cline{3-5} \cline{7-10}
layer & basis set & \multicolumn{3}{c}{Full basis} & & \multicolumn{4}{c}{Perturbation theory} \\
\hline 
1 & $6s5p4d$ 		& 73	& 425	& 4\,537	&& 41 & (16)	& 247	& 4\,538 \\ 
2 & $6s5p5d4f$ 	& 335	& 3\,267	& 4\,868	&& 188 & (74)	& 1\,897	& 4\,868 \\ 
3 & $7s6p5d4f$ 	& 749	& 6\,915	& 4\,930	&& 421 & (166)	& 4\,019	& 4\,930 \\
4 & $7s6p6d5f$ 	& 1\,315	& 13\,761	& 4\,880	&& 740 & (292)	& 7\,993	& 4\,880 \\
5 & $8s7p6d5f$ 	& 2\,033	& 20\,425	& 4\,888	&& 1\,145 & (452)	& 11\,869	& 5\,102 \\
6 & $8s7p7d6f$ 	& 2\,903	& 31\,275	& 4\,859	&& 1\,636 & (646)	& 18\,167	& 5\,069 \\
7 & $9s8p7d6f$ 	& 3\,925	& 40\,955	& 4\,858	&& 2\,213 & (874)	& 23\,797	& 5\,279 \\
8 & $9s8p8d7f$ 	& 5\,099	& 55\,809	& 4\,842	&& 2\,876 &(1\,537)	& 32\,419	& 5\,268 \\
9 & $10s9p8d7f$ 	& 6\,425	& 68\,505	& 4\,848	&& 3\,625& (2\,286)	& 39\,803	& 5\,495 \\
10 & $10s9p9d8f$~~	& --	& --	& -- 	&& 4\,460&(3\,121)	& 50\,749	& 5\,508 \\
\hline \hline
\end{tabular}
\end{center}
\end{table}

\begin{figure}[t]
\begin{center}
\includegraphics[width=0.84\textwidth]{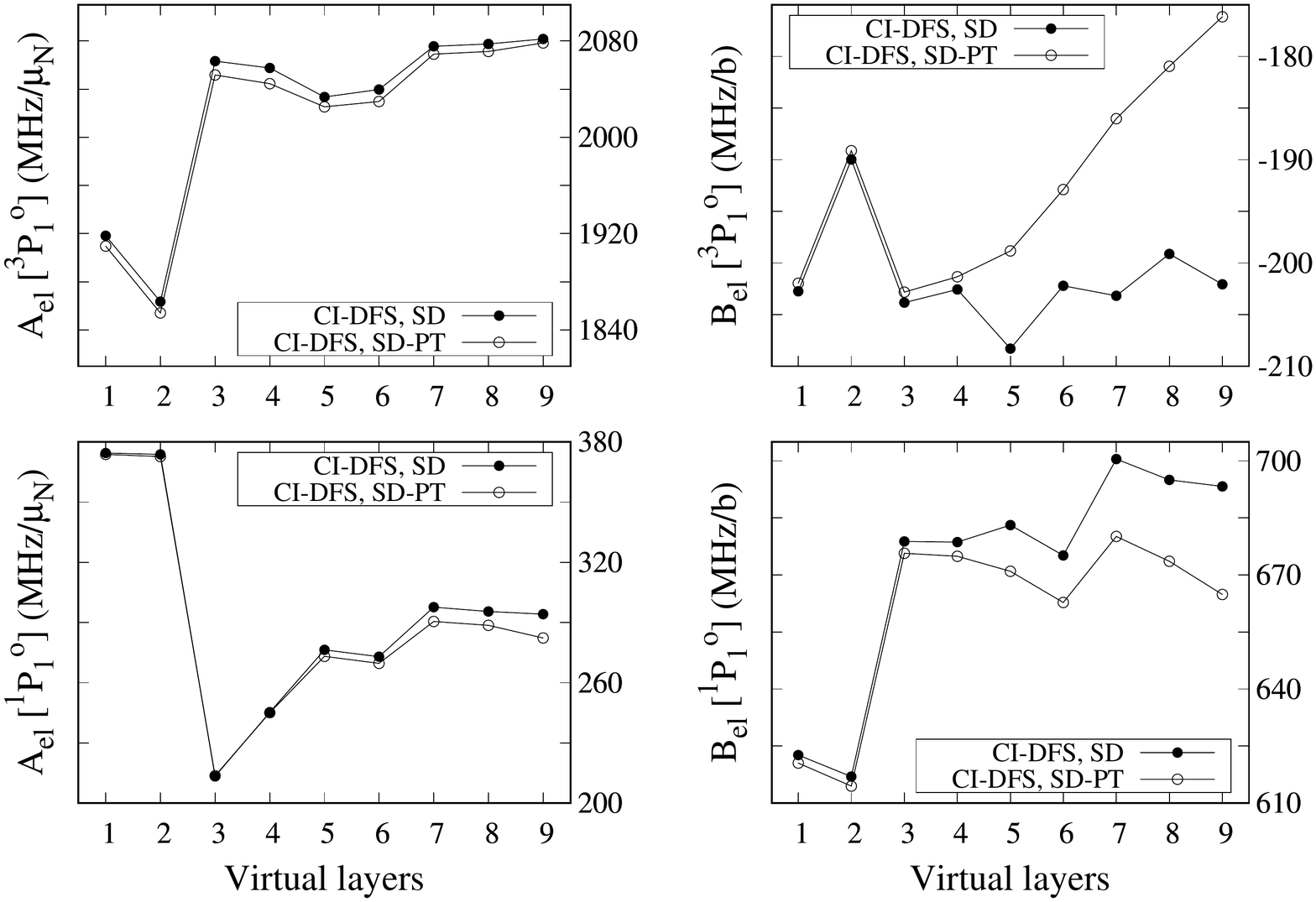} 
	\caption {The convergence of the electronic hyperfine factors $A_{\rm el} [{}^3\!P_1^{\rm o}]$ , $A_{\rm el} [{}^1\!P_1^{\rm o}]$ (left panels) and $B_{\rm el} [{}^3\!P_1^{\rm o}]$, $B_{\rm el} [{}^1\!P_1^{\rm o}]$
	(right panels) with the increasing number of virtual orbital layers, optimized by employing the 
	CI-DFS method. The solid circles represent the results from the direct(full basis) calculations, while the empty circles illustrate the values obtained using perturbation theory (PT). The numbers of virtual orbital layers on the x-axes are equivalent to the numbers given in column~1 of Table~\ref{Tab:NO_configurations}. 
	For more details, see text in Secs.~\ref{subsection.CIDFS} and~\ref{section.CIDFS_calculations}.} \label{Fig:NO_results}
\end{center}
\end{figure}

For each virtual orbital layer that was used in the CI-DFS calculations, the corresponding orbital basis set, numbers of configurations and computed energy separations $\Delta E$ between the targeted $^3\!P_1^{\rm o}$ and $^1\!P_1^{\rm o}$ states are listed in Table~\ref{Tab:NO_configurations}. One can see that the resulting energy difference $\Delta E = E[{}^1\!P_1^{\rm o}]-E[{}^3\!P_1^{\rm o}]$ from the direct (full basis) calculations is well converged, in contrast to the PT calculations, where the $\Delta E$ value is not saturated.
By constructing the reduced one-particle density matrix, one can calculate the isotope-independent M1 and E2 hyperfine splitting constants, respectively, given by Eqs.~\eqref{A_el} and \eqref{B_el}. 
For both non-PT (solid circles) and PT (empty circles) bases, the convergence patterns of the isotope-independent hyperfine constants $A_{\text{el}}[{}^{3}\!P^{\rm o}_{1}]$, $A_{\text{el}}[{}^{1}\!P^{\rm o}_{1}]$, $B_{\text{el}}[{}^{3}\!P^{\rm o}_{1}]$, and $B_{\text{el}}[{}^{1}\!P^{\rm o}_{1}]$ are shown in Fig.~\ref{Fig:NO_results}. It is seen in the figure that the electronic E2 hyperfine factors $B_{\rm el}$ are more sensitive to variations of the orbital basis set, in comparison to the electronic M1 hyperfine factors $A_{\rm el}$, and for that reason, their theoretical uncertainties are larger. In addition, we observe that the results from the non-PT and PT calculations progressively diverge as the number of virtual PT orbitals increases. That being so, and taking also into account the weaker stability of the PT energy separation value, the results from the perturbative treatment can only be used for estimating the theoretical error bars (see Sec.~\ref{section.ErrorBudget}), and not for extending the basis further.
The final results of the CI-DFS calculations are based on the largest non-PT orbital basis set (corresponding to the $10s9p8d7f$ set of orbitals) and are shown below:
%
\begin{equation}\label{eq:cidfs_res}
\begin{split}
A_{\rm el}\left[^3\!P^{\rm o}_1\right]&= 2\ 082~{\rm MHz}/\mu_N  \quad 
B_{\rm el}\left[^3\!P^{\rm o}_1\right]=-202~ {\rm MHz/b}\,;\\
A_{\rm el}\left[^1\!P^{\rm o}_1\right]&=294~{\rm MHz}/\mu_N    \quad 
B_{\rm el}\left[^1\!P^{\rm o}_1\right]= 693~{\rm MHz/b}\,.
\end{split}
\end{equation}


\section{Final value and evaluation of accuracy}
\label{section.ErrorBudget}
In Secs.~\ref{section.Heidelberg}--\ref{section.CIDFS_calculations}, four different computational approaches for evaluating the electronic hyperfine factors $A_{\text{el}}$ and $B_{\text{el}}$ of the $^{1,3}\!P^{\rm o}_1$ excited states in neutral tin were presented. In this section, we solely focus on the $B_{\text{el}}[^{1}\!P^{\rm o}_1]$ value that can be used to extract the quadrupole moments, $Q$, of tin isotopes for which spectroscopic data are available. The four independent sets of calculations yielded four individual values of $B_{\text{el}} [^1\!P^{\rm o}_{1}]$. 
In summary, we obtained
$B_{\text{el}}[^{1}\!P^{\rm o}_1]$~=~622~MHz/b from the S-MR-MCDHF calculations (see Eq.~\eqref{eq:zoltan_res} in Sec.~\ref{section.Heidelberg}),
$B_{\text{el}}[^{1}\!P^{\rm o}_1]$~=~778~MHz/b from the SrD-SR-MCDHF calculations
(see Eq.~\eqref{eq:result778} in Sec.~\ref{section.SRcalculations}), $B_{\text{el}}[^{1}\!P^{\rm o}_1]$~=~717~MHz/b from the SrD-MR-MCDHF calculations (see Table~\ref{table:sensitivity} in Sec.~\ref{section.MRcalculations}), and $B_{\text{el}}[^{1}\!P^{\rm o}_1]$~=~693~MHz/b from the CI-DFS calculations (see Eq.~\eqref{eq:cidfs_res} in  Sec.~\ref{section.CIDFS_calculations}). We ultimately arrive at $B_{\text{el}}[^{1}\!P^{\rm o}_1]$~=~703~MHz/b by taking the average of these values. 

As a crude estimate of the uncertainty of the concluding $B_{\text{el}}[^{1}\!P^{\rm o}_1]$ value, we can consider the half-range of the aforementioned individual results, i.e., 78~MHz/b. Yet, if one wants to be in a position to discuss atomic, or nuclear, properties and their underlying physics, a rigorous assessment of the uncertainties of the computed values is required. In recent years, atomic physicists have been putting great efforts into providing reliable uncertainties on their theoretical results~\cite{SafJoh:2008a,SafSaf:2011a,Chuetal:2016a,Bieron:zinc4s4p:2018,Wanetal:2018d}. In line with these efforts, the following subsections take into account a number of considerations to determine the accuracy of the final $B_{\text{el}}[^{1}\!P^{\rm o}_1]$ value. Some of these considerations are only applicable to one (or more) particular set(s) of calculations (see Sec.~\ref{subsection.model-specific-errors}), while others are analogously applied to all four separate results (see Sec.~\ref{subsection.A_theor-A_expt}). Statistical principles are implemented (see Sec.~\ref{subsection.standard_dev}), and former outcomes from computations of electronic hyperfine factors, regarding atomic states with electronic structure similar to the structure of the $5s^25p6s\,\,^{1,3} \! P^{\rm o}_{1}$ states in Sn~I, are also used as an estimate of the accuracy of the $B_{\text{el}}[^{1}\!P^{\rm o}_1]$ value deduced in this work (see Sec.~\ref{subsection.ZnAnalogy}).

\subsection{Model-specific uncertainties}
\label{subsection.model-specific-errors}

In each of the four independent approaches that were discussed in the previous sections, the wave functions (radial orbitals and configuration mixing coefficients) that describe the atomic states were obtained based on various approximations with respect to the orbital basis and the list of configuration states. In Sec.~\ref{section.sensitivity}, the sensitivity of the SrD-SR-MCDHF+RCI and SrD-MR-MCDHF+RCI results to the orbital basis was investigated by combining the radial orbital basis obtained in one of these two computational approaches with the CSF expansions used in the RCI computations of the other approach. As seen in Table~\ref{table:sensitivity}, these combinations gave rise to $B_{\text{el}}[^{1}\!P^{\rm o}_1]$ values that range from 709~MHz to 760~MHz/b. 
The half-range of these values yields an uncertainty of 26~Hz/b.
Further, in the CI-DFS calculations, the electronic hyperfine factors were computed using both non-PT and PT orbital bases. The comparison between the non-PT and PT results for different orbital basis sets suggests an uncertainty of 70~MHz/b in the deduced value of $B_{\text{el}}[^{1}\!P^{\rm o}_1]$.
Lastly, in the instance of the SrD-SR-MCHDF calculations, the outcome for the $B_{\text{el}}[^{1}\!P^{\rm o}_1]$ value is the average of four separate values. One can, thus, assume an error bar corresponding to the half-range of these values, i.e., 20~MHz/b.

\subsection{Difference between the theoretical and experimental $A_{\rm el}$ values}
\label{subsection.A_theor-A_expt}

The deviation of the calculated 
M1 hyperfine constant from the experimental value, A$_{\rm theor}$ - A$_{\rm expt}$, is often assumed to be a measure of the overall accuracy of the hyperfine structure calculations~\cite{BieronAu2009,Bieron:zinc4s4p:2018,Schetal:2020a}. In Sec.~\ref{section.Estimates}, the experimental A$_{\rm expt}[^3\!P^{\rm o}_1]$ value was used to accordingly shift the resulting $B_{\text{el}}[^1\!P^{\rm o}_1]$ values from all three  approximations, i.e., SrD, SD, and SDT, that were used in the SrD-SR-MCDHF+RCI calculations. Considering only the $B_{\text{el}}[^1\!P^{\rm o}_1]$ result from the most extensive SDT calculation and evaluating the difference $B_{\rm el}(\text{SDT})-B_{\rm el}(\text{SDT})^{\rm shifted}$ yields an error estimate of 32~MHz/b. When applying this shift to the final results of the remaining calculations, we acquire three more error bars; 54~MHz/b from the S-MR-MCDHF calculations, 64~MHz/b from the SrD-MR-MCDHF calculations, and 91~MHz/b from the CI-DFS calculations.


\subsection{Statistical standard deviation}
\label{subsection.standard_dev}
The individual results provided by the four independent sets of calculations
could be regarded as a statistical sample 
and, in that case, the average value $\mu$ and the standard deviation $\sigma$ can be evaluated. For the $\{622,\  693, \ 717,\  778\}$ set of $B_{\text{el}}[^1\!P^{\rm o}_1]$-values, it is $\mu\pm\sigma=703\pm56$~MHz/b, which positions $B_{\text{el}}[^1\!P^{\rm o}_1]$ between 591~MHz/b and 815~MHz/b within the 2$\sigma$ condition ($95\%$). In this manner, we obtain another uncertainty estimate equivalent to $56$~MHz.
\subsection{Zinc analogy}
\label{subsection.ZnAnalogy}

In a recent paper~\cite{Bieron:zinc4s4p:2018}, the quadrupole moment $Q(^{67}{\rm Zn})$ was evaluated based on $11$ independent multiconfiguration calculations of the EFG $\propto B_{\rm el}$ for the $4s 4p \,\, ^3 \! P^{\rm o}_{1}$ and $4s 4p \,\, ^3 \! P^{\rm o}_{2}$ states in Zn~I. The final accuracy of the calculated EFGs was estimated using the scatter of the individual results of these 11 calculations, resulting in a relative error of about $8\%$. The valence structure of the $4s 4p \,\, ^3 \! P^{\rm o}_{1,2}$ states in neutral zinc is quite similar to the structure of the $5s^25p 6s \,\, ^{1,3} \! P^{\rm o}_{1}$ tin states, which are of interest in this work; in both cases, there are two electrons outside the closed shells, and the orbitals of these valence electrons have similar angular symmetries. Thereby, one expects that, for atomic calculations using similar computational approaches, the relative error bars of the computed hyperfine factors will be comparable. Adopting the $8\%$ relative error bar, an uncertainty of $56$~MHz/b is inferred for the $B_{\text{el}}[{}^1\! P^{\rm o}_{1}]$ value deduced in the present paper.


\subsection{Final accuracy}
\label{section.Final}

The considerations above lead to diverse error bars, which, according to the order of their appearance in the text, are (in units of MHz/b): $78, \ 26, \ 70, \ 20, \ 32, \ 54, \ 64, \ 91, \ 56$, and $56$. The largest of these uncertainties, i.e., $91$~MHz/b, places $B_{\text{el}}[{}^1\! P^{\rm o}_{1}]$ between 521~MHz/b and 885~MHz/b within the $2\sigma$ condition, which is a rather conservative choice. On the other hand, the smallest of all these error estimates, i.e., 20~MHz/b, positions $B_{\text{el}}[{}^1\! P^{\rm o}_{1}]$ between 663~MHz/b and 743~MHz/b within the $2\sigma$ condition. This interval does not overlap with all individual $B_{\text{el}}[{}^1\! P^{\rm o}_{1}]$-values resulting from the four independent sets of calculations and therefore, such an error bar is not appropriate. Assuming that some of the obtained error bars possibly overestimate the uncertainty in our concluding $B_{\text{el}}[{}^1\! P^{\rm o}_{1}]$ value, and that a few others might underestimate it, we believe that the rounded value of $50$~MHz/b is
a reasonable choice.
The final result of the present paper, then, becomes 
\begin{equation}
B_{\text{el}}[{}^1\! P^{\rm o}_{1}]~=~703~\pm~50~\text{MHz/b} \ ,
\end{equation}
localizing $B_{\text{el}}[{}^1\! P^{\rm o}_{1}]$ between 603 MHz/b and 803 MHz/b with $95\%$ confidence.
Comparing the value above with the $B_{\text{el}}[{}^1\! P^{\rm o}_{1}]$ value resulting from the simplest DHF calculation yields a difference of $\sim17\%$, which is significantly higher than the ($50/703 \simeq$) $7\%$  uncertainty claimed in this work.

\section{Quadrupole moments}
\label{section.Quadrupoles}
The computed $B_{\text{el}}[{}^1\! P^{\rm o}_{1}] \propto$ EFG value can be used to deduce the nuclear quadrupole moments $Q(^{A}\text{Sn})=B/B_{\text{el}}$ of the tin isotopes for which the E2 hyperfine constant $B$ was measured. The $B_{\text{el}}$ value resulting from the present multiconfiguration calculations was recently employed by Yordanov \textit{et~al.}~\cite{Yordanov:20} to extract the nuclear quadrupole moments of odd-$A$ tin isotopes.
%
As mentioned in the introduction, the final value of the EFG for the $5s^25p6s~^1\!P_1^o$ state has been slightly shifted 
from 706(50)~MHz/b that was reported and used in Ref.~\cite{Yordanov:20} to 703(50)~MHz/b in the present paper.
The $Q$-values listed in the last column of Table~1 in Ref.~\cite{Yordanov:20} should, therefore, be increased by a factor of 706/703.

For a few tin isotopes, more than one experimental values of E2 hyperfine constants are available, allowing us to compare the extracted quadrupole moments. The E2 hyperfine constant $B$[$^1\!P^{\rm o}_1$] for the $I=5/2$ state of $^{109}$Sn was measured independently in Refs.~\cite{Ebeetal:87a,Yordanov:20}, and their results are, respectively, $B[^1\!P^{\rm o}_1]=212.0(27.0)$~MHz and $B[^1\!P^{\rm o}_1]=154(5)$~MHz.  
In Ref.~\cite{Ebeetal:87a}, the computed $B_{\text{el}}[^1\!P^{\rm o}_1]=596$~MHz/b value is also available, despite the fact that they used the data related to the $^3\!P^{\rm o}_1$ state, i.e., $B[^3\!P^{\rm o}_1]=-43.0(12.0)$~MHz and $B_{\text{el}}[^3\!P^{\rm o}_1]=-138$~MHz/b, to extract the quadrupole moment $Q(^{109}$Sn$)=310(100)$~mb.
By combining the experimental $B[^1\!P^{\rm o}_1]$ result of Ref.~\cite{Yordanov:20} with the presently computed $B_{\text{el}}[^1\!P^{\rm o}_1]=703(50)$, we obtain $Q(^{109}$Sn$)=219(7)(16)$, which significantly differs from the above-mentioned value of $Q(^{109}$Sn$)=310(100)$ mb. These two quadrupole moments merely overlap with each other due to the large uncertainty of 100~mb in the latter value. Further, the $B_{\text{el}}[^1\!P^{\rm o}_1]$ value given in Ref.~\cite{Ebeetal:87a} barely overlaps with the present $B_{\text{el}}[^1\!P^{\rm o}_1]$ value, which
strengthens the need for re-computing the electronic hyperfine factors by using state-of-the-art programs and performing large-scale calculations.

Finally, taking the $^{119}$Sn isotope as an example, we propose 
\begin{equation}
Q(^{119}{\rm Sn}) = -0.176(4)(12)~{\mbox{b}} \; ,
\end{equation}
where (12) represents the theoretical uncertainty of 7\% of the $B_{\text{el}}[^1\!P^{\rm o}_1]$ deduced in this work and (4) represents the experimental uncertainty of the measured $B[^1\!P^{\rm o}_1]$ in Ref.~\cite{Yordanov:20}.
We notice that the theoretical uncertainty suggested above is about three times larger than the experimental uncertainty. In the previous section, a number of considerations was taken into account to provide a well-grounded estimate of the theoretical uncertainties in our final $B_{\text{el}}[^1\!P^{\rm o}_1]$ value. Nonetheless, one should always remain cautious towards error estimates of electronic hyperfine factors deduced from atomic calculations and of the corresponding error bars in the evaluated nuclear quadrupole moments $Q$. The element bismuth is a good example of the difficulties in estimating such error bars.
Table~I in Ref.~\cite{BieronBi2001} lists the proposed values of the nuclear quadrupole moment $Q$ for the $^{209}$Bi isotope. The error bars of several of those values not only do not overlap, but they do not even touch each other (to make all of the error bars overlap, the relative uncertainties would have to exceed 50\%). We should, however, also note here that the valence structure of the tin atom is less complicated and less demanding computationally than the valence structure of the bismuth atom, and we are confident enough that the deduced error bars in this paper are trustworthy. 
%
\section{Conclusions}
\label{section.Conclusions}
We presented the details of the theoretical calculations of the isotope-independent electric quadrupole hyperfine constant $B_{\text{el}}$ $\propto$ EFG, which was recently used to extract nuclear quadrupole moments, $Q$, of tin isotopes~\cite{Yordanov:20}. Four independent computational approaches were employed to finally provide the
value of $B_{\text{el}}$~=~703(50)~MHz/b for the $5s^25p6s$ $^1\!P^{\rm o}_1$ excited state of Sn~I. Three of these approaches were based on the variational MCDHF method as implemented in the {\sc Grasp} packages, whilst the fourth approach relied on the CI-DFS theory. 
The convergence of $B_{\text{el}}[{}^1\! P^{\rm o}_{1}]$ was monitored as the CSF expansions were enlarged by allowing single, double, and, depending on the correlation model, also triple, electron substitutions from the reference configuration(s). Efforts were made to provide a realistic theoretical uncertainty for the final $B_{\text{el}}[{}^1\! P^{\rm o}_{1}]$ value by accounting for statistical principles, the correlation with the isotope-independent magnetic dipole hyperfine constant $A_{\text{el}}$, and previous calculations of electronic hyperfine factors on systems with electronic structure similar to that of Sn~I.

The deduced relative accuracy of the present atomic {\it ab initio} calculations of the EFGs is of the order of $7\%$, leading to even larger uncertainties in the extracted $Q$(Sn)-values due to the uncertainty in the measured $B$. This level of accuracy is certainly inferior to the deduced $Q$(Sn)-values from the solid-state density functional calculations performed by Barone \textit{et~al.}~\cite{Barone:08}, which are about an order of magnitude more accurate. In general, the accuracy of the atomic {\it ab initio} calculations of EFGs strongly depends on the valence structure of the atom, or ion, in question.
We should note that, in the extreme case of lithium-like systems, the relative uncertainties of the atomic calculations of hyperfine structures can be limited to $0.001 - 0.01\%$~\cite{Tong1993,Yan1996,BieronLi1996,BieronBeF1999,Yerokhin2008a,Yerokhin2008b}.
%
Even though the tin atom is far more demanding computationally than the lithium-like systems, an atomic calculation of hyperfine structures
with lower uncertainty would be possible for singly-ionized tin, with one electron outside closed shells, and it would be even more accurate, for triply-ionized tin, which has one electron outside the $n<5$ core.
Such calculations, as the latter, would likely challenge the accuracy of the solid-state methods.
We hereby encourage experimentalists to consider one, or both, of the above-mentioned ions.

Interestingly, we observed that all computed $A_{\rm el}[^{3}\!P_{1}^{\rm o}]$ values are always smaller than the experimental $A_{\rm el}^{\rm exp}[^{3}\!P_{1}^{\rm o}]=2\ 398$~MHz/$\mu_{\rm N}$ value, independently of the computational method or the correlation model. This could be explained by the lack of variational freedom intrinsic to the layer-by-layer optimization strategy, which hinders the contraction of spectroscopic orbitals when CV correlation is accounted for. In the specific case of tin, the spectroscopic $4d$ soft shell, i.e., lying between the core and the valence orbitals, is expected to be highly sensitive to CV correlation that might not be effectively captured. Natural orbitals were recently used, as an efficient tool to overcome the limitation of the layer-by-layer optimization scheme, to estimate hyperfine structure constants in Na~I. Thanks to the radial re-organization of the orbitals, the spectroscopic orbitals are ultimately contracted, which affects both the magnetic dipole and electric quadrupole electronic hyperfine factors~\cite{Schetal:2020a}. Further investigations on the usefulness of the natural orbitals in the calculations of hyperfine structures are in progress.
\begin{acknowledgments}
\noindent 

SS is a FRIA grantee of the Fonds de la Recherche Scientifique$-$FNRS. MG acknowledges support from the FWO \& FNRS Excellence of Science Programme (EOS-O022818F), PJ acknowledges support from the Swedish Research Council (VR) under contract 2015-04842, and IIT acknowledges support by the RFBR Grant No. 18-03-01220.
\end{acknowledgments}

\clearpage
\appendix
\section{Sensitivity of the hyperfine constants and stability of the EFGs ratio} 

\label{section.MCDHFBP}
The extreme sensitivity of $A_{\text{el}}[{}^1\!P_{1}^{\rm o}]$ to correlation models is not really surprising if one performs calculations using the quasi-relativistic 
Hartree-Fock+Breit-Pauli~\cite{FBJbook} 
method in the single configuration approximation. In the Breit-Pauli (BP) scheme, the low value of the ratio $A_{\text{el}}[^1\!P_{1}^{\rm o}]$/$A_{\text{el}}[^3\!P_{1}^{\rm o}]$ can indeed be understood. The $A_{\text{el}}$ value of the {\it pure} $^3\!P_1^{\rm o}$ i.e., without considering any relativistic $LS$-term mixing, arises  from the addition of the three contributions~\cite{Jonetal:93a}, orbital, spin-dipole and contact term, which interfere positively. 
On the other hand, the $A_{\text{el}}$ value of the {\it pure} $^1\!P_1^{\rm o}$ is only made of a (larger) orbital contribution, the total spin value ($S=0$) forbidding the two other contributions. For $J=1$, the two singlet and triplet symmetries are mixed with relative phases that result from the orthogonality constraints. The eigenvector dominated by the triplet character has the same signs of both components,
which makes the $A_{\text{el}}$ value even larger  than the one of the pure triplet (increase of 40\%). For the state dominated by the singlet, strong cancellation occurs due to the triplet contamination, reducing the $A_{\text{el}}$ value by 61\%. Strong cancellation in the estimation of a property usually involves high uncertainty.  \\

The ``sharing rule''~\cite{BauCha:76a,Ishetal:97a} that is used to quantify configuration mixing from the measured isotope shifts can be applied to the term-mixing analysis of the electric field gradients. In the single-configuration approximation, the ratio EFG[$^3\!P_{1}^{\rm o}]/$EFG$[^{1}\!P_{1}^{\rm o}$] is exactly $-1/2=-0.5$, resulting from angular momentum algebra, when using the same orbital sets for describing both levels.  Assuming a simple $\; ^3\!P^{\rm o}- \; ^1\!P^{\rm o}$ mixing for $J=1$, we have
\begin{equation}
\Psi(`` \;  ^3\!P^{\rm o}_1 \; ") = a \vert ^3\!P^{\rm o}_1 \rangle + b \vert ^1\!P^{\rm o}_1 \rangle \; , 
\end{equation}
\[ \Psi(`` \;  ^1\!P^{\rm o}_1 \; ") = b \vert ^3\!P^{\rm o}_1 \rangle - a \vert ^1\!P^{\rm o}_1 \rangle \; ,\]
where $\ket{^{3,1}\!P^{\rm o}_1}$ are the two lowest $J^\Pi=1^-$ states resulting from pure $LS$-terms and $\Psi(`` \, ^{3,1}\!P^{\rm o}_1 \, ")$ are the corresponding mixed states.  Using the analytical ratio EFG[$^1\!P_{1}^{\rm o}]/$EFG$[^3\!P_{1}^{\rm o}]= -2$, one can estimate
\begin{equation}
\mbox{EFG}[``\; ^3\!P^{\rm o}_1 \; "] = \mbox{EFG}[\; ^3\!P^{\rm o}_1] ( a^2 - 2b^2) \; ,
\end{equation} 
\[ \mbox{EFG}[``\; ^1\!P^{\rm o}_1 \; "] = \mbox{EFG}[\; ^3\!P^{\rm o}_1] ( b^2 - 2a^2) \; , \]
from which we deduce
\begin{equation}
 R = 
\mbox{EFG}[`` \; ^3\!P^{\rm o}_{1} \; "] / \mbox{EFG}[`` \; ^1\!P^{\rm o}_{1} \; "] = \frac{a^2 - 2b^2}{b^2 - 2a^2} \; .
\end{equation}
Adopting for this ratio a reasonable guess that is guided by the experimental result from Ref.~\cite{Yordanov:20} and that offers numerical simplicity,  $R = -1/4 $, 
one gets the following analytical eigenvector compositions
\begin{equation}
\Psi(``\; ^3\!P^{\rm o}_1 \; ") =    \frac{\sqrt{7}}{3} \;  \vert ^3\!P^{\rm o}_1 \rangle +   \frac{\sqrt{2}}{3} \; \vert ^1\!P^{\rm o}_1 
 \rangle \; ,
 \end{equation}
\[ \Psi(``\; ^1\!P^{\rm o}_1 \; ") =   \frac{\sqrt{2}}{3} \;  \vert ^3\!P^{\rm o}_1 \rangle  - \frac{\sqrt{7}}{3}  \; \vert ^1\!P^{\rm o}_1 
 \rangle \; . \]
In other terms, the ratio $\mbox{EFG}[`` \, ^3\!P^{\rm o}_{1} \, "] / \mbox{EFG}[`` \, ^1\!P^{\rm o}_{1} \, "]$  only reflects the singlet-triplet mixing in this simple model. We should not be surprised by its relative stability for more elaborate models.
Extracting the $^3\!P^{\rm o}$ character $(a^2)$ from the lowest ($`` \, ^3\!P^{\rm o}_{1} \, "$) BP eigenvector obtained in a simple MR model mixing the $\{5s^2 5p6s,\ 5s5p5d6s, \ 5p^3 6s \}$ configurations, we get after renormalization $a^2=0.77617$ from which we determine $\mbox{EFG}[`` \, ^3\!P^{\rm o}_{1} \, "] / \mbox{EFG}[`` \, ^1\!P^{\rm o}_{1} \, "]=-0.247$ according to
\begin{equation}
\mbox{EFG}[`` \, ^3\!P^{\rm o}_{1} \, "] / \mbox{EFG}[`` \, ^3\!P^{\rm o}_{1} \, "] = \frac{a^2 - 2b^2}{b^2 - 2a^2} 
= -\frac{3a^2-2}{3a^2-1}  \; .
\end{equation}
For the other BP eigenvector ($`` \, ^1\!P^{\rm o}_{1} \, "$), we have $a^2= 0.77641$ from which one confirms the ratio
$\mbox{EFG}[`` \, ^3\!P^{\rm o}_{1} \, "] / \mbox{EFG}[`` \, ^1\!P^{\rm o}_{1} \, "]=-0.248$. The latter value is not too far from the above $R = -1/4$ ratio value that would be obtained from the hypothetical $(a^2 = 7/9 ; b^2 = 2/9)$ singlet-triplet mixing, taking into account that 
(i) one trusts the nonrelativistic ratio EFG[$^3\!P_{1}^{\rm o}]/$EFG$[^1\!P_{1}^{\rm o}$]~$=-1/2$  of the single-configuration approximation, (ii) the BP eigenvector has to be renormalized, 
and (iii) one assumes no contamination by other $LS$ symmetries ($\; ^3\!D^{\rm o}_1, \; ^5\!P^{\rm o}_1,\; ^5\!D^{\rm o}_1,\; ^5\!F^{\rm o}_1, \;^7\!D^{\rm o}_1, \ldots, \; ^{25}\!X^{\rm o}_1$). 



%

\end{document}